\documentstyle[11pt]{article}

\newlength{\dinwidth}
\newlength{\dinmargin}
\def\slepton{\widetilde \ell}

\def\slr{\slepton_{R}}
\def\sll{\slepton_{L}}
\def\squark{\widetilde q}
\def\msl{m_{\slepton}}
\def\msq{m_{\squark}}
\def\msll{m_{\sll}}
\def\mslr{m_{\slr}}
\def\photino{\widetilde \gamma}
\def\zino{{\widetilde{Z}}}
\def\wino{{\widetilde{W}}}
\def\sfermion{\widetilde f}

\def\gluino{\widetilde g}

\def\sneutrino{\widetilde \nu}
\def\sele{\widetilde e}
\def\ser{\sele_{R}}

\def\sql{\squark_{L}}

\def\msf{m_{\sfermion}}
\def\msn{m_{\sneutrino}}
\def\msg{m_{\gluino}}

\def\stau{\widetilde \tau}
\def\staul{\stau_{L}}
\def\staur{\stau_{R}}
\def\su{\widetilde{u}}
\def\sd{\widetilde{d}}
\def\sc{\widetilde{c}}
\def\ss{\widetilde{s}}
\def\sul{\su_{L}}
\def\sdl{\sd_{L}}
\def\sur{\su_{R}}
\def\sdr{\sd_{R}}
\def\sclr{\sc_{L,R}}
\def\sslr{\ss_{L,R}}
\def\sulr{\su_{L,R}}
\def\sdlr{\sd_{L,R}}
\def\msul{m_{\sul}}
\def\msur{m_{\sur}}
\def\msdl{m_{\sdl}}
\def\msdr{m_{\sdr}}
\def\msclr{m_{\sclr}}
\def\msulr{m_{\sulr}}
\def\msdlr{m_{\sdlr}}
\def\msslr{m_{\sslr}}
\def\mch{m_{H^{+}}}
\def\mA{m_{A}}
\def\mH{m_{H}}
\def\st{\widetilde{t}}
\def\sb{\widetilde{b}}

\def\sz1{{\widetilde{Z}}_{1}}
\def\szs{{\widetilde{Z}}_{2}}
\def\szt{{\widetilde{Z}}_{3}}
\def\szf{{\widetilde{Z}}_{4}}
\def\szk{{\widetilde{Z}}_{k}}
\def\swl{{\widetilde{W}}_{1}}
\def\swh{{\widetilde{W}}_{2}}

\def\msz1{m_{\sz1}}
\def\mszs{m_{\szs}}
\def\mszt{m_{\szt}}
\def\mszf{m_{\szf}}
\def\mswl{m_{\swl}}
\def\mswh{m_{\swh}}

\def\gev{{\rm GeV}}

\def\rs{{\sqrt{s}}}
\def\tanbe{\tan\beta}
\def\nle{{\stackrel{<}{\sim}}}
\def\nge{{\stackrel{>}{\sim}}}
\def\goto{\rightarrow}

\def\mt{m_{t}}
\def\mb{m_{b}}
\def\mh{m_{h}}
\def\stl{\st_{1}}
\def\stls{\st^{*}_{1}}
\def\sth{\st_{2}}
\def\sbl{\sb_{1}}
\def\sbr{\sb_{R}}
\def\sbh{\sb_{2}}
\def\mst{m_{\st}}
\def\mstl{m_{\stl}}
\def\msth{m_{\sth}}
\def\msb{m_{\sb}}
\def\msbl{m_{\sbl}}
\def\msbh{m_{\sbh}}
\def\mz{m_{Z}}
\def\mw{m_{W}}
\def\tht{\theta_{t}}

\def\tew{\theta_{W}}
\def\cbar{\overline{c}}
\def\sw{\sin^{2}\theta_{W}}
\def\cw{\cos^{2}\theta_{W}}
\def\muo{\mu_{\infty}}
\def\mgo{M_{\infty}}
\def\afo{A_{\infty}}
\def\mfo{m_{\infty}}
\def\mgut{M_{X}}
\def\mtil{\widetilde{m}}
\def\Xtil{\widetilde{X}}
\def\misEt{\slash\hspace{-8pt}E_{T}}
\def\pos{\slash\hspace{-6pt}p_{1}}
\def\pts{\slash\hspace{-6pt}p_{2}}
\def\prs{\slash\hspace{-6pt}p_{3}}
\def\mot{m_{12}^{2}}
\def\mor{m_{13}^{2}}
\def\mtr{m_{23}^{2}}

\def\mssqh{\mtil^{2}_{Q_{3}}}

\def\mssuh{\mtil^{2}_{U_{3}}}

\def\msh1{\mtil^{2}_{H_{1}}}
\def\msh2{\mtil^{2}_{H_{2}}}
\def\Ktil{\widetilde{K}}
\def\Itil{\widetilde{I}}
\begin{document}
{}~~~\\
\vspace{10mm}
\begin{flushright}
ITP-SU-94/02 \\
RUP-94-06
\end{flushright}
\begin{center}
  \begin{Large}
   \begin{bf}
A Light Scalar Top Quark in MSGUT
 \\
   \end{bf}
  \end{Large}
  \vspace{5mm}
  \begin{large}
    Tadashi Kon
\footnote{
e-mail address : d34477@jpnac.bitnet, mtsk@jpnkektr.bitnet
}
and Toshihiko Nonaka $^{\dagger}$
\footnote{
e-mail address : rra001c@jpnrky00.bitnet
}
\\
  \end{large}
\vspace{3mm}
Faculty of Engineering, Seikei University, Tokyo 180, Japan \\
$^{\dagger}$Department of Physics, Rikkyo University, Tokyo 171, Japan \\
\vspace{5mm}
\end{center}
\vskip50pt
\begin{quotation}
\noindent
\begin{center}
{\bf Abstract}
\end{center}
We reexamine a possibility for the existence of
a light supersymmetric partner of the top quark (stop)
with mass 15$\sim$16GeV in the framework of
the minimal supergravity GUT model (MSGUT).
Such light stop could explain the slight excess of
the high $p_{T}$ cross section of the $D^{*\pm}$-meson
production in the two-photon process at TRISTAN.
We find two types of solution for the RGEs in the MSGUT
allowing the existence of the light stop.
The type~{\uppercase\expandafter{\romannumeral 1}}
[type~{\uppercase\expandafter{\romannumeral 2}}] solution is
characterized by the heavy [light] top quark,
$\mt\simeq$ 150GeV [100GeV], and
the light [heavy] squarks, $\msq\simeq$ 150GeV [300GeV].
It is found that the type~{\uppercase\expandafter{\romannumeral 2}}
solution is  more favorable because these parameter sets seem to
satisfy all constraints settled by the recent collider experiments
as well as by the non-accelerator observations and
the cosmological considerations.
We point out that the existence of such a stop could change the dominant
decay mode of sparticles, the top quark and the Higgs bosons.
Consequently,the present experimental bounds on the supersymmetric
parameters as well as on masses of the top and the Higgs could be
weakened substantially.
However, the allowed parameter region is rather restricted
and in turn masses and mixing parameters of
the other SUSY partners as well as masses of the Higgs and
the top are severely constrained.
For example,
$75\gev\nle\msg\nle85\gev$, $\mswl\nle50\gev$,
$\msl\simeq\msq\simeq300\gev$,
$\tht\simeq0.9$,
$\mh\nle65\gev$ and $90\gev\nle\mt\nle100\gev$.
\end{quotation}
\vfill\eject
\section{\it Introduction}
In spite of much efforts of many experimental colleagues the nature
has not yet opened "the new physics windows" and
the standard model (SM) of the electroweak interaction
has been established precisely except for
the lack of the top quark and the Higgs boson.
The mass of top quark is now expected to be larger than about 100GeV
and the other quarks and the leptons are known to be
much lighter than the top \cite{top}.
One of the models beyond the SM,
the supersymmetric (SUSY) standard models
\cite{Nilles}, draw much attention since
they can solve the naturalness problem in the SM and can achieve
naturally the unification of gauge interactions at the GUT
scale \cite{running}.
It is known that the naturalness criterion in the SUSY models
inevitablely predicts the existence of many SUSY partners
(sparticles) with masses below the order of magnitude of one TeV.
Therefore, the search for the sparticles
is one of the most important purposes of
the present and future collider experiments.
Up to now, there is no clear evidence for the existence of the sparticles
at the present collider experiments
and the lower mass bounds for them are growing up day by day.
Should we wait for LEP-{\uppercase\expandafter{\romannumeral 2}},
LHC or some next linear colliders to open the SUSY window?

In the minimal SUSY standard model (MSSM)
the slepton $\slepton$, the lighter chargino $\swl$
or the scalar top quark (stop) $\stl$ is expected to be
the lightest charged sparticle .
While the model independent lower bounds
on masses of the sleptons and the chargino
have been settled by LEP , $\msl$ , $\mswl$ $>$ $\mz/2$,
there is a window allowing the existence of
a very light stop,
$\mstl$ $<$ $\mz/2$, even if we have
no evidence at LEP \cite{DH} and Tevatron \cite{Baer}.
{}From a theoretical point of view, moreover, it is argued that
one of stops $\stl$ is naturally
lighter than the other squarks and the sleptons
because of the large top mass \cite{stop,HK}.
Such a possibility is very interesting
because this is the SUSY counterpart of
the fact that the leptons and the other quarks are
much lighter than the top.
We should notice such a possibility seriously because the stop could
be discovered first at the collider experiments even before the top.

Enomoto et al. in the TOPAZ group at TRISTAN
have reported a slight excess of the high $p_T$
cross section of $D^{*\pm}$-meson production
in a two-photon process \cite{Enomoto}.
The disagreement between the measured value and the standard
model prediction now becomes  $3.5\sigma$ level \cite{EnomotoT},
which should be compared
to $1.5\sigma$ reported previously \cite{Enomoto}.
Although there remains a possibility that such a excess could be
explained by the large contribution of the gluon structure
function of the photon \cite{resolve},
there is another exciting way to interpret this enhancement, i.e.,
it is the signature of the pair production of the stop with
mass $\nle$ $20\gev$.
Since such a light stop will decay into the charm-quark plus the
lightest neutralino \cite{HK},
the signature of the stop production will be the
charmed meson production with large missing energies.
This signature would  resemble  the charmed-hadron production
in the two-photon process at $e^+e^-$ colliders.
Enomoto et al. have pointed out that the stop with mass about
$15\sim16\gev$ and the neutralino with mass about
$13\sim14\gev$ could well explain the
experimental data.

In previous papers \cite{ours},
we have investigated the possibility
for the existence of the stop
with mass 15$\sim$16GeV in the framework of
the minimal supergravity GUT model (MSGUT) \cite{sugra}.
We have pointed out that
the existence of such a stop could change the dominant
decay mode of some particles, especially that of the gluino,
and could weaken the present experimental bound on the SUSY parameters.
We have found the light stop solutions for the renormalization
group equations (RGEs) \cite{RGE} and
such solutions inevitably set constraints on
masses and mixing parameters of
the other SUSY partners as well as masses of the Higgs and
the top, for example,
$\msg\nle85\gev$, $\mswl\nle55\gev$, $110\gev\nle\msl\nle140\gev$,
$120\gev\nle\msq\nle160\gev$,
$\tht\simeq0.9$, $\mh\nle65\gev$ and
$\mt\nle135\gev$.

In this paper we give a new set of solutions for the RGEs
to allow the light stop $\mstl\simeq15\gev$ called
type~{\uppercase\expandafter{\romannumeral 2}} solution
in addition to the solution obtained previously
(type~{\uppercase\expandafter{\romannumeral 1}}),
where we also present our calculational scheme in detail.
The type~{\uppercase\expandafter{\romannumeral 2}} solution
is characterized by the heavy squarks and sleptons,
$\msl$ $\simeq$ $\msq$ $\simeq$ $300\gev$,
and the light top, $\mt$ $\nle$ $100\gev$.
We show this type of solution is more favorable
if we consider the recent collider data as well as the
nucleon decay and the dark matter constraints.

The paper is  organized as follows.
The theoretical bases in the MSSM for lightness of the stop are
summarized in Sec.2.
In Sec.3 we  reconsider the present bounds on the stop mass and
those on the gaugino parameters.
The two types of solution for the RGEs in the MSGUT are given in Sec.4.
In Sec.5 we choose favorable solution by a consideration of
some experimental informations.
Some phenomenological implications of the light stop
are disscussed in Sec.6.
Sec.7 is  devoted to summary and conclusion.

\section{\it Light stop : its theoretical bases}

In the framework of the MSSM \cite{Nilles},
the stop mass matrix in the ($\st_{L}$, $\st_{R}$) basis is expressed by
\begin{equation}\renewcommand{\arraystretch}{1.3}
{\cal M}^{2}_{\st}=\left(
                 \begin{array}{cc}
                   m^{2}_{\st_{L}} & a_{t}m_{t} \\
                   a_{t}m_{t} & m^{2}_{\st_{R}}
                 \end{array}
                \right),
\label{matrix}
\end{equation}
where $\mt$ is the top mass.
The SUSY mass parameters $m_{\st_{L, R}}$ and $a_{t}$
are parametrized in the following way \cite{susy} :
\begin{eqnarray}
m^{2}_{\st_{L}}&=&{\widetilde{m}}^{2}_{Q_{3}}
  +\mz^{2}\cos{2\beta}\left({\frac{1}{2}}-
{\frac{2}{3}}\sin^{2}\tew\right)+m^{2}_{t}, \label{mstl}\\
m^{2}_{\st_{R}}&=&{\widetilde{m}}^{2}_{U_{3}}
  +{\frac{2}{3}}\mz^{2}\cos{2\beta}\sin^{2}\tew+m^{2}_{t}, \label{mstr}\\
a_{t}&=&A_{t}+\mu\cot\beta,
\end{eqnarray}
where $\tanbe$, $\mu$ and $A_{t}$ denote
the ratio of two Higgs
vacuum expectation values ($=v_{2}/v_{1}$), the SUSY Higgs mass
parameter and
the trilinear coupling constant, respectively.
The soft breaking masses of the third generation doublet
${\widetilde{m}}_{Q_{3}}$ and the up-type singlet
${\widetilde{m}}_{U_{3}}$ squarks are related to those
of the first (and second) generation squarks as
\begin{eqnarray}
{\widetilde{m}}^{2}_{Q_{3}}&=&{\widetilde{m}}^{2}_{Q_{1}}
-{\widetilde{I}}, \label{mq3}\\
{\widetilde{m}}^{2}_{U_{3}}&=&{\widetilde{m}}^{2}_{U_{1}}
-2{\widetilde{I}}, \label{mu3}
\end{eqnarray}
where ${\widetilde{I}}$ is a function proportional to
the top quark Yukawa coupling $\alpha_{t}$ and
is determined by the RGEs in the MSGUT
(see Sec.4 and Appendices A and B).
Throughout of this paper we adopt the notation of Ref.\cite{Hikasa}.

There are two origins for lightness of the stop compared to
the other squarks and sleptons,
{\romannumeral 1}) smallness of the diagonal soft
masses $m^{2}_{\st_{L}}$ and $m^{2}_{\st_{R}}$ and
{\romannumeral 2}) the left-right stop mixing.
Both effects are originated from
the large Yukawa interaction of the top.
The origin {\romannumeral 1}) can be easily seen from
Eqs.(\ref{mstl})$\sim$(\ref{mu3}).
The diagonal mass parameters $m^{2}_{\st_{L}}$ and $m^{2}_{\st_{R}}$
in Eq.(\ref{matrix}) have possibly small values
owing to the negative large
contributions of ${\widetilde{I}}$ proportional to $\alpha_{t}$ in
Eqs.(\ref{mq3}) and (\ref{mu3}).
It should be noted that this contribution is also important in
the radiative SU(2)$\times$U(1) breaking in the MSGUT.
The Higgs mass squared has an expression similar to
Eqs.(\ref{mq3}) and (\ref{mu3}) ;
\begin{equation}
{\widetilde{m}}^{2}_{H_{2}}=
{\widetilde{m}}^{2}_{L_{1}}
-3{\widetilde{I}},
\label{Higgsmass}
\end{equation}
where ${\widetilde{m}}^{2}_{L_{1}}$ denotes
the soft breaking mass of the first generation doublet slepton.
The large contribution of ${\widetilde{I}}$ enables
${\widetilde{m}}^{2}_{H_{2}}$
to become negative at an appropriate weak energy scale.
In order to see another origin {\romannumeral 2}) we should diagonalize
the mass matrix ({\ref{matrix}).
The mass eigenvalues are obtained by
\begin{equation}
m^{2}_{\stl\atop\sth}
         ={\frac{1}{2}}\left[ m^{2}_{\st_{L}}+m^{2}_{\st_{R}}
             \mp \left( (m^{2}_{\st_{L}}-m^{2}_{\st_{R}})^{2}
            +(2a_{t}m_{t})^{2}\right)^{1/2}\right].
\label{stopmass}
\end{equation}
and the corresponding mass eigenstates are expressed by
\begin{equation}
\left({\stl\atop\sth}\right)=
\left(
{\st_{L}\,\cos\tht-\st_{R}\,\sin\tht}
\atop
{\st_{L}\,\sin\tht+\st_{R}\,\cos\tht}
\right),
\end{equation}
where $\tht$ denotes the mixing angle of stops :
\begin{eqnarray}
\tan\tht=
{\frac{a_{t}\,m_{t}}
  {m^{2}_{\st_{R}}-\mstl^{2}}}.
\label{tantht}
\end{eqnarray}
We see that if the SUSY mass parameters and the top
mass are of the same order of magnitude,
small $\mstl$ is possible owing to the cancellation
in the expression (\ref{stopmass}) \cite{stop,HK}.

After the mass diagonalization
we can obtain the interaction Lagrangian for the
mass eigenstate $\stl$.
We note, in particular, that the stop coupling to
the $Z$-boson ($\stl\stls Z$) depends sensitively on
the mixing angle $\tht$.
More specifically, it is proportional to
\begin{equation}
C_{\stl}\equiv {\frac{2}{3}}\sin^{2}\tew - {\frac{1}{2}}\cos^{2}\tht.
\label{c}
\end{equation}
Note that for the special value of $\tht$$\sim$0.98,
the $Z$-boson coupling completely vanishes \cite{DH}.

\section{\it Present bounds on stop mass and gaugino parameters}
\subsection{\it stop mass bounds}

Before a discussion of experimental bounds on the stop mass $\mstl$,
we examine the decay modes of the stop.
In the MSSM, the stop
lighter than the other squarks and gluino
can decay into the various final states :
$\stl$ $\to$ $t\,\sz1$ (a),
$ b\swl$ (b),
$ b\ell\sneutrino $ (c),
$ b\nu\slepton $ (d),
$ bW\sz1 $ (e),
$ bff'\sz1 $ (f),
$ c\sz1$ (g),
where $\sz1$, $\swl$, $\sneutrino$ and $\slepton$, respectively, denote
the lightest neutralino, the lighter chargino, the sneutrino and the
charged slepton.
If we consider the light stop with mass smaller than 20GeV,
the first five decay modes (a) to (e) are kinematically
forbidden due to the model independent
lower mass bounds for respective particles ; $m_{t}$$\nge$60GeV
\renewcommand{\thefootnote}{\fnsymbol{footnote}}
\footnote{
This is the bound independent on the decay mode of the top.
In fact, the CDF bound, $\mt>110\gev$,
will not be applicable in the case of the light stop,
as we will discuss in Sec.6  },
$m_{\swl}$$\nge$45GeV, $m_{\slepton}$$\nge$45GeV and
$m_{\sneutrino}$$\nge$40GeV.
So (f) and (g) left .
Hikasa and Kobayashi \cite{HK} have shown that
the one-loop mode $\stl\to c\sz1$ (g) dominates over the
four-body mode $\stl\to bff'\sz1$ (f).
This is absolutely true in the case considered here, because
the mode (f) is negligible by the kinematical suppression,
$\mstl \sim \msz1 +\mb$.
So we can conclude that such a
light stop will decay into
the charm quark jet plus the missing momentum taken away
by the neutralino with  almost 100$\%$ branching ratio.
Note that the width of stop in this case is very small,
i.e., of the order of  magnitude of eV.

Naively, it will be expected that
Tevatron and/or LEP can set severe
bounds on the stop mass through the processes ;
$gg$ $\to$ $\stl\stl^{*}$ $\to $ $c\cbar\sz1\sz1$
(Tevatron) and/or
$Z$ $\to$ $\stl\stl^{*}$ (LEP).
However, the situation is not so obvious.
Baer et al. \cite{Baer} have performed the analyses of the
experimental data of 4$pb^{-1}$ integrated luminosity
Tevatron running,
and have obtained the results that the stop could easily escape
detection if $m_{\sz1}$ $\nge$ 10GeV.
Such a large neutralino mass could make the charm quark jets softer.
Consequently the stop production cross section plotted against
the missing transverse energy becomes smaller than
the present upper bounds,
where they impose cuts on the missing transverse energy
\footnote{
Recent analyses has been reported in Ref.\cite{Tata}.
}.
Moreover, we should point out that LEP cannot exclude a light stop
$\mstl<\mz/2$ by measuring the extra width of the $Z$-boson.
In fact we find that there is no bound on the stop mass if the
mixing angle $\tht$ is larger than about 0.7,
where we take the experimental limit
$\Delta\Gamma_{Z}<28$MeV (95\% C.L.) \cite{ours}.
The origin of such a sensitivity of $\Gamma(Z\to\stl\stl^{*})$
is in the fact that the $\stl\stls Z$ coupling is proportional to
$C_{\stl}$ (\ref{c}) \cite{DH}.

Okada \cite{bsg} has investigated
possible bounds on masses of the stop and the neutralino
from the experimental data of the $b\goto s\gamma$ decay.
He has shown that the light stop with mass $\mstl\nle$20GeV
has not been excluded by the data.
It has been also pointed out by Fukugita et al. \cite{FMYY} that
the existence of the light stop does not conflict with the
experimental bounds on $\Delta\rho$ and $K^{0}$ $-$ $\bar{K}^{0}$
mixing.

Recently, the stringest limit comes from the direct
searches for the stop at $e^+e^-$ colliders.
New analyses of the direct search by the VENUS group
at TRISTAN show that {\it the TOPAZ stop}
($\mstl=15\sim16$GeV and $\msz1=13\sim14$GeV)
just confronts the experimental bounds \cite{VENUST}.
In fact it seems that such a stop has been excluded for
$\mstl-\msz1>3\gev$.
Moreover, the OPAL group have reported the bounds on the
($\mstl$, $\msz1$, $\tht$) plane from the direct searches
at LEP \cite{Kobayashi}.
They have shown the light stop $\mstl<\mz/2$ only
survives if $\mstl-\msz1<2.2\gev$ and $1.15>\tht>0.85$.
A similar bound has been reported by the DELPHI group \cite{DELPHI}.

We should say, after all,
{\it the TOPAZ stop} has not been excluded
only if $\mstl-\msz1<2.2\gev$ and $1.15>\tht>0.85$.
In the following, we examine such a possibility in the framework of
the MSGUT and investigate its phenomenological implications.

\subsection{\it bounds on gaugino parameters}

In the MSSM, masses and mixing parameters of the gaugino-higgsino
sector are determined by three parameters
$\mu$, $\tanbe$ and $M_{2}$, where $M_2$ denotes the soft breaking
SU(2) gaugino mass.
Some regions in the ($\mu$, $\tanbe$, $M_2$) parameter space
have already excluded
by the negative searches for the SUSY particles at
some collider experiments.
First, we consider the experimental data at LEP ;
lower bound on the mass of lighter chargino,
\begin{equation}
\mswl\nge45\gev,
\label{chargino}
\end{equation}
upper bound on the branching ratio of
the visible neutralino modes of the $Z$ \cite{DELPHI,visZ},
\begin{equation}
{\rm BR}(Z\goto vis.)\equiv
\sum_{{i,j}\atop{i=j\neq 1}}\Gamma(Z\goto\zino_{i}\zino_{j})/
\Gamma_{Z}^{\rm tot} < 5\times 10^{-6} ,
\label{visible}
\end{equation}
and upper bound on the invisible width of the $Z$ \cite{L3},
\begin{equation}
\Gamma(Z\goto\zino_{1}\zino_{1}) < 16.2{\rm MeV} .
\label{invisible}
\end{equation}
  Next we should discuss bounds on the gaugino parameters from
the hadron collider experiments.
If we assume {\it the GUT relation},
\begin{equation}
\msg = M_{3} = {\frac{\alpha_{s}}{\alpha}}\sw M_{2}
\label{GUTrel}
\end{equation}
in the MSGUT, the gluino mass $\msg$ bounds from the hadron colliders
could be converted into the bounds on $M_{2}$ \cite{Hidaka}.
Naively accepted gluino mass bound at CDF is
\begin{equation}
\msg\nge 150\gev \qquad\quad {\rm (90\% C.L.)},
\label{sgboundd}
\end{equation}
which can be easily converted into the bound on $M_2$
by Eq.(\ref{GUTrel}) as $M_2$ $\nge$ $44\gev$.
(Note that {\it the GUT relation} (\ref{GUTrel}) depends sensitively
on $\sw$ and $\alpha_{s}$. Here we take $\sw=0.232$ and
$\alpha_{s}=0.113$.)

In Fig.1 we show the region excluded by the experimental data
(\ref{chargino}), (\ref{visible}), (\ref{invisible}) and
(\ref{sgboundd}) in the ($\mu$, $M_2$) plane for $\tanbe=$2,
where the regions inside of each contour have been excluded.
We also plot a contour of $\msz1=13\sim14$GeV which can explain the
TRISTAN data as mentioned above.
At first sight, the neutralino with mass $13\sim14$GeV has already
been excluded by the bounds from LEP and Tevatron experiments.
(Note that the contour of $\msz1=13\sim14$GeV lies also
in the excluded region for $\mu>0$.)
We see that the stringest bound comes from the constraint on
the visible width of the $Z$-boson at LEP (\ref{visible}) [B]
or from the gluino search at CDF (\ref{sgboundd}) [D].
Fortunately, however, these bounds are rather model dependent.
As we will show in Sec.5, the existence of the light stop can
weaken both bounds substantially.
Here we should adopt the model independent bounds from
the lower mass bound on the lighter chargino (\ref{chargino}) [A]
and upper bound on
the invisible width of the $Z$-boson (\ref{invisible}) [C].
Then we find the allowed region,
$-170\gev$ $\nle$ $\mu$ $\nle$ $-20\gev$ for $\tanbe$ $=$ $2$.
If we take larger (smaller) values of $\tanbe$, the allowed
region becomes narrower (wider).
We find that the allowed region disappears
for $\tanbe\nge 2.8$.
Furthermore, we see that $\msz1=13\sim14$GeV corresponds to
$M_{2}=22\sim24$GeV in the allowed region and
we can find that this correspondence is
independent of the values of $\tanbe$.
Consequently, we can take $M_{2}=22\sim24$GeV as an input value in
the following calculation.
The allowed region in the ($\mu$, $\tanbe$) plane fixed by $M_{2}$ $=$
$22\gev$ or  $24\gev$ is shown in Fig.2.
It is worth mentioning that the lightest neutralino $\sz1$ is
almost the photino $\photino$ in the allowed parameter range
in Fig.2.
In fact, the photino component of the neutralino is larger than
$98$\% in the range.

\section{\it MSGUT analysis}
\subsection{\it solutions for RGEs}

Before presenting our results for analysis,
we summarize briefly the calculational scheme in the MSGUT
\cite{Hikasa}.
All physical parameters go from the GUT scale
$\mgut$ down to low energies
governed by the RGEs \cite{RGE}.
All RGEs for the MSGUT parameters are presented in Appendix A.
Here we neglect all Yukawa couplings except for the top.
This is not a bad approximation as long as $\tanbe$ is not too large
($\ll\mt /\mb$), which is the case we consider here,
$\tanbe\nle 2.8$.
In order to solve the RGEs we must fix the boundary conditions
for independent parameters.
At $\mgut$ we set the following conditions on the parameters ;
the SUSY Higgs mass parameter
$\mu(\mgut)=\muo$ and three soft breaking mass parameters :
the common scalar mass
$\mtil_{\sfermion}^{2}(\mgut)$ $=$ $\mtil_{H_{i}}^{2}(\mgut)$ $=$
$\mfo^{2}$, the gaugino mass
$f_{3}^{-1}M_{3}(\mgut)$ $=$ $f_{2}^{-1}M_{2}(\mgut)$ $=$
$f_{1}^{-1}M_{1}(\mgut)$ $=$ $\mgo$ and
the trilinear coupling
$A_{\tau}(\mgut)$ $=$ $A_{b}(\mgut)$ $=$ $A_{t}(\mgut)$ $=$ $\cdots$
$=$ $\afo$. As usual, we take the Higgs mixing parameter $B$ as
$B(\mgut)=\afo-\mfo$.
We present the analytical formulae for solutions of the RGEs in the
general case $f_{i}\neq 1$ in Appendix B.
Hereafter we take {\it the GUT relation} (\ref{GUTrel}) i.e.,
$f_{1}=f_{2}=f_{3}=1$, and discuss briefly the case $f_3\neq1$
in Appendix C.

As for the evolution of the gauge couplings $\alpha_{i}(t)$
and the gaugino masses $M_{i}(t)$, we take the input values
$\alpha^{-1}(\mz)=128.8$ and $\sw=0.232$.
Assuming that the SUSY scale is not too different from $\mz$,
we may use the SUSY beta function at all scales above $\mz$
for simplification.
Then one finds $\mgut=2.1\times 10^{16}\gev$, $\alpha_{\infty}^{-1}$
($=\alpha_{3}^{-1}(\mgut)$ $=\alpha_{2}^{-1}(\mgut)$
$=\alpha_{1}^{-1}(\mgut)$) $=24.6$ and
$\alpha_{3}(\mz)=0.113$.

All physics at weak scale $\mz$ are determined by the six
parameters ($\mfo$, $\afo$, $\mgo$, $\mu$, $\tanbe$, $\mt$).
There are, moreover, two conditions imposed on the parameters
to have the correct scale of SU(2)$\times$U(1) breaking.
So we can reduce the number of the independent parameters
to four out of the six.
Here we take the four independent input parameters as
($\mgo$, $\mu$, $\tanbe$, $\mt$).
As we have discussed in Sec.3.2, furthermore,
we can fix one of the input values,
$M_{2}=22\sim24\gev$, which corresponds to $\mgo=26.7\sim29.1\gev$
for $\sw=0.232$.
After all, there remain the only three arbitrary parameters
($\mu$, $\tanbe$, $\mt$).

\subsection{\it two types of solutions}

We seek numerical solutions for the RGEs to give a
light stop with mass $\mstl=15\sim16\gev$ varying the three
parameters ($\mu$, $\tanbe$, $\mt$).
Then we find two types of solution,
type~{\uppercase\expandafter{\romannumeral 1}} and
type~{\uppercase\expandafter{\romannumeral 2}}.
The type~{\uppercase\expandafter{\romannumeral 1}} solution
has already been presented at the previous papers \cite{ours}.
Common properties of both solutions are
{\romannumeral 1})
mass of lighter chargino $\swl$ is smaller than $55\gev$,
{\romannumeral 2})
the lighter CP even neutral Higgs boson is relatively light,
$\mh\nle 65\gev$ and
{\romannumeral 3})
the stop mixing angle $\tht$ is severely
limited as $\tht$ $\simeq$ 0.9.
It is interesting that $\tht\simeq0.9$ is not {\it input}
but {\it output} of the MSGUT calculation.
In fact this value can save the light stop from the recent bounds
from the direct searches at LEP \cite{Kobayashi,DELPHI}
as discussed in Sec.3.1.
The differences between the two types of solution are in
masses of the top quark and the squarks (sleptons)
in the first and the second generations.
The relatively heavy [light] top and the light [heavy] squarks
(sleptons) characterizes  the
type~{\uppercase\expandafter{\romannumeral 1}}
[type~{\uppercase\expandafter{\romannumeral 2}}] solution.
In the next section, we consider
some experimental informations
in order to choose favorable solution and
to get further constraints  on the parameters.

\section{\it Choice of parameter sets and further constraits}

\subsection{\it gluino searches at CDF}

It is known that the bound (\ref{sgboundd})
from the negative gluino search at CDF is not realistic.
To get realistic bound we must include the cascade decays in the
analyses \cite{cascade}.
The gluino mass bound at CDF taken into account of the cascade decays
$\gluino\goto q\overline{q}\zino_{2,3,4}$ and
$\gluino\goto ud\wino_{1,2}$ as well as the direct decay
$\gluino\goto q\overline{q}\zino_{1}$ has been reported as \cite{CDF}
\begin{equation}
\msg\nge 95\gev \qquad\quad {\rm (90\% C.L.)}
\label{sgboundc}
\end{equation}
for $\mu=-250\gev$ and $\tanbe=2$, for example.
We should note, however, that the bound (\ref{sgboundc})
has been obtained based on the assumption that
$\mstl>\msg$ and the gluino can not decay into the stop.
This is not the case we consider here.
In fact, the gluino can decay into the stop pairs,
$\gluino\goto\stl\stls\sz1$, which becomes another seed
for the cascade decay
because the stop and neutralino could be light enough.

The Feynman diagrams for the gluino decay
are depicted in Fig.3 and the analytical formula for the decay width is
presented in Appendix D.
In Fig.4 we show the $\msg$ dependence of the branching ratio of
gluino, where we include the mode $\gluino\goto\stl\stls\sz1$
and sum up quark flavors $q, q' = u, d, c, s$.
We take $\tanbe=2.0$, $\mu=-150\gev$, $\mstl=15\gev$,
$\tht=0.9$, and $M_{2}=22\gev$,
and take $\msg$ as a free parameter.
The squark masses and the top mass are taken as
($\msq$, $\mt$) $=$ ($2\msg$, $130\gev$) for (a) and
($3.8\msg$, $95\gev$) for (b),
where $\msq$ $\equiv\msulr$ $=\msdlr$ $=\msclr$ $=\msslr$.
Note that Fig.4 (a) and (b) respectively correspond to the
solutions type~{\uppercase\expandafter{\romannumeral 1}} and
type~{\uppercase\expandafter{\romannumeral 2}}.
The branching ratio of the direct decay mode
$\gluino\goto q\overline{q}\zino_{1}$, which is important in the
$\gluino$ search in terms of large $\misEt$
signature, is reduced substantially
even for the light gluino with mass $\msg\nge60\gev$ .
Therefore, we should reconsider the UA2 bound $\msg\nge79\gev$
\cite{UA2} obtained under the assumption
BR$(\gluino\goto q\overline{q}\zino_{1})$ $\sim$ 100\%
as well as the CDF bound (\ref{sgboundc}).
For the value $\msg =75\gev$ determined by
{\it the GUT relation},
BR$(\gluino\goto q\overline{q}\zino_{1})$ $\sim$ 20\% [0.6\%]
for the type~{\uppercase\expandafter{\romannumeral 1}}
[type~{\uppercase\expandafter{\romannumeral 2}}] solution,
which should be compared with
BR$(\gluino\goto q\overline{q}\zino_{1})$ $\sim$ 70\%
obtained when there is no stop mode.
We can find that if we take larger values of $\msq$
and/or smaller values of $\mt$,
BR$(\gluino\goto q\overline{q}\zino_{1})$ is reduced rapidly.
In this case the Tevatron bound (\ref{sgboundc}) would be diminished
significantly.
This is because the width of the stop mode
$\Gamma(\gluino\goto\stl\stls\sz1)$
becomes larger for smaller values of $\mt$ and
all the other widths become smaller for larger values of $\msq$,
as we can see from Figs.3 and 4.

We try to simulate the Monte-Carlo calculation in order to get
the gluino mass bounds from the CDF gluino searches.
In Fig.5 we show the expected number of events in
4.3$pb^{-1}$ integrated luminosity Tevatron running.
In this calculation, we take following kinematical selection cuts
\cite{cascade}.
(i) We coalesce partons within
$\Delta R$ $=$ $\sqrt{(\Delta\eta)^2+(\Delta\phi)^2}$ $<$ $0.7$
into single jets.
We also require that all jets satisfy $|\eta_{ij}|$ $<$ $3.5$,
and each jet must have $E_{T}$ $>$ $15\gev$.
The highest $E_{T}$ cluster is also required to be central
($|\eta|<1$).
(ii) We require that there be no jet with $E_{T}$ $>$ $5\gev$
within a $30^{\circ}$ cone back to back in azimuth with the
leading jet.
(iii) We require $\misEt$ $>$ $40\gev$.
We can find that
the lower mass bounds for gluino
are respectively about $95\gev$ and $60\gev$
for type~{\uppercase\expandafter{\romannumeral 1}} and
type~{\uppercase\expandafter{\romannumeral 2}} solution.

\subsection{\it second neutralino searches at LEP}

In Sec.3 we have noted that the second neutralino search
through measuring the visible width of the $Z$-boson
could set severe constraints on the MSSM basic parameters
($M_2$, $\tanbe$, $\mu$).
The experimental upper bound reported recently is
(\ref{visible}) \cite{DELPHI,visZ} and this bound can be
converted to the excluded region in the
($M_2$, $\mu$) plane fixed $\tanbe$ $=$ $2$ as shown in Fig.1.
We should note that the leptons or quarks (jets) signatures
have been used in seraching for the second neutralino
through the direct decay $\szs$ $\to$ $f\bar{f}\sz1$.
The bound (\ref{visible}) has been obtained under
the assumption of BR($\szs$ $\to$ $f\bar{f}\sz1$) $=$ $100$\%.
Here again we should consider the appearance of the light stops
in the final state of the second neutralino decay.
In fact, if the mass of second neutralino $\mszs$ is larger than
$2\mstl+\msz1$ $\simeq$ $43\gev$, it can decay into the stops.

The Feynman diagrams for the second neutralino decay
are depicted in Fig.6 and the analytical formula for the decay width is
presented in Appendix D, where we neglect a contribution
Fig.6(4) because $\sz1\simeq\photino$ {\it and}
the stop coupling to the $Z$-boson (\ref{c}) is expected to be
rather small in our case.
In Fig.7 we show the $\mt$ dependence of the branching ratio of
second neutalino.
We take $\tanbe=2.0$, $\mu=-150\gev$, $\mstl=15\gev$,
$\tht=0.9$, $\mh=60\gev$, $\alpha=-0.6$ and $M_{2}=22\gev$,
where $\alpha$ denotes the Higgs mixing angle \cite{GH}.
The squark masses and the trilinear coupling $A_{t}$ are taken as
($\msq$, $A_t$) $=$ ($2\msg$, $300\gev$) in (a) and
($3.8\msg$, $600\gev$) in (b),
where we take $\msl$ $=$ $\msq$ for simplicity.
It is worth mentioning that
$\Gamma(\szs\to\stl\stls\sz1)$ dominates over
$\Gamma(\szs\to f\bar{f}\sz1)$ for large $\mt$
because the virtual Higgs exchange diagram Fig.6(3) gives
large contribution to the stop width.
The predicted values in the parameter sets for
type~{\uppercase\expandafter{\romannumeral 1}} and
type~{\uppercase\expandafter{\romannumeral 2}}
can be read from Fig.7 (a) with $\mt\simeq130\gev$
and (b) with $\mt\simeq95\gev$, respectively.
The branching ratio of the direct decay mode
$\szs\goto f\overline{f}\sz1$, which is important in the
$\szs$ search in terms of the leptons or quarks (jets)
signature, is reduced substantially as
BR$(\szs\goto f\overline{f}\sz1)$ $\simeq$ 25\% [2.5\%]
for type~{\uppercase\expandafter{\romannumeral 1}}
[type~{\uppercase\expandafter{\romannumeral 2}}] solution.
Therefore, we should reconsider the LEP bound
(\ref{visible}) \cite{DELPHI,visZ} obtained under the assumption
BR$(\szs\goto f\overline{f}\sz1)$ $\sim$ 100\%.
Simply we can obtain the effective bounds on the observed branching
ratio as
\begin{equation}
{\rm BR}(Z\goto vis.)_{\rm obs}^{\rm upper} =
{\frac{{\rm BR}(Z\goto vis.)_{\rm exp}^{\rm upper}}
{{\rm BR}(\szs\goto f\overline{f}\sz1)}},
\end{equation}
and we get its  numerical value, $2\times10^{-5}$ [$2\times10^{-4}$]
for the type~{\uppercase\expandafter{\romannumeral 1}}
[type~{\uppercase\expandafter{\romannumeral 2}}] solution.

After all, we obtain the correct excluded region
\footnote {Rigorously speaking, the contour of $\msg$
depends slightly on $\mu$.
In Fig.8 we neglect such a sensitivity since our conclusion
is not affected by it.}
in the ($M_2$, $\mu$)
plane fixed $\tanbe=2$
for type~{\uppercase\expandafter{\romannumeral 1}} and
type~{\uppercase\expandafter{\romannumeral 2}} solutions
depicted in Fig.8 (a) and (b), respectively.
We realize that the neutralino with mass $13\sim14$GeV is still excluded
for the type~{\uppercase\expandafter{\romannumeral 1}} solution.
However, a window ,
$-170\gev$ $\nle$ $\mu$ $\nle$ $-70\gev$, which allows the existence
of the light neutralino, opens up for the
type~{\uppercase\expandafter{\romannumeral 2}} solution.

\subsection{\it proton decay constraint}

Besides the collider experiments we should consider the constraints
on the model parameters from the non-accelerator obsevations
and the cosmological considerations.
In the MSGUT there are the dangerous dimension five operators
which contribute to the fast nucleon decay \cite{pdecay}.
The nucleon decay life time $\tau_{N}$ in the framework of the
SU(5) SUSY GUT model has been investigated in detail and
the present upper bounds on $\tau_{N}$ have been
converted to the constraints on the masses of the chargino and
the sfermions.
The most stringest constraint comes from the non-observation for
the decay $N\goto K{\nu}_{\mu}$.
Hisano et al. have shown that the squarks and the sleptons with masses
$200\gev$ have been excluded for $\mswl$ $\nle$ $50\gev$ even if we
take the most conservative parameters \cite{pdecay}.
This result seems to indicate that
the type~{\uppercase\expandafter{\romannumeral 2}} solution
is  more favorable.

\subsection{\it dark matter candidate}

It is expected that the LSP in the SUSY model can be a good
candidate for the dark matter of the universe \cite{DarkM}.
In our case the LSP is the lightest neutralino with mass
$13\gev$ $\sim$ $14\gev$ and it is almost the photino.
The relic abandance $\Omega h^{2}$
of the light photino has been calculated
and the upper bound on the sfermion masses can be obtained
if we consider the universe will not be overclosed
$\Omega h^{2}$ $<$ $1$.
Naively, the photino with mass $\nle$ $20\gev$ can be a
dark matter candidate only when $\msq$, $\msl$ $\nle$ $160\gev$.
At first insight,
the type~{\uppercase\expandafter{\romannumeral 1}} solution
is  more favorable and this fact seems to contradict with
the result from the nucleon decay
constraints discussed in the last subsection.

Recently, however, Fukugita et al. \cite{FMYY} have pointed out that
the co-annihilation process
$\photino\photino\to gg$ can be occured through
the chain process $\photino + c \to \st$ and
$\st\st^{*}\to gg$ in the case of the light stop.
The cross section of such a process is enhanced when
$m_{\st}$ and $m_{\photino}$ are close to degenerate.
They have shown that the cosmological limit
$\Omega h^{2}$ $>$ $1$ can be avoided
even for $\msq$, $\msl$ $\simeq$ $300\gev$
in so far as $m_{\st}-m_{\photino}$ $\nle$ $5\gev$.
Note that the type II solution gives $\msq$, $\msl$ $\simeq$ $300\gev$
and the type~{\uppercase\expandafter{\romannumeral 2}} solution
can survive.
It has been pointed out that the proton decay favors a large
value of $\xi_{0}\equiv\mfo/\mgo$ but the cosmology of
the neutralino dark matter disfavors large value of
$\xi_{0}$ \cite{DarkM}.
So the type~{\uppercase\expandafter{\romannumeral 2}} solution
satisfies both constraints.

\subsection{\it neutral Higgs searches at LEP}

A result from the negative searches for the MSSM Higgs at
LEP could set another constraint on the SUSY parameter space.
The present limit on the MSSM (SM like) Higgs mass is
\begin{equation}
\mh\nge 61.5\gev \qquad ({\rm 95\% C.L.})
\label{higgsbnd}
\end{equation}
for $\sin(\beta-\alpha)\simeq 1$ \cite{Kobayashi}.
Note, however, that this bound has been obtained
based upon the assumption that
the Higgs boson does not decay into the stop.
Here we must consider the fact that
the neutral Higgs could have
a dominant decay mode $h\goto \stl\stls$ with almost 100\%
branching ratio if the stop is light enough.
In this case energies of visible jets from the Higgs production
would become softer and it can be smaller than the detection lower cuts.
Therefore, if we incorporate such a decay mode in data analyses,
the present lower bounds are expected to be weakened.

We try to simulate the Monte-Carlo calculation in order to get
the Higgs mass bounds from the LEP Higgs searches.
In Fig.9 we show the expected number of events in
93.5 $pb^{-1}$ integrated luminosity LEP running.
Here we assume the SM Higgs will decay into $b\bar{b}$ with
100\% branching ratio for simplicity.
It is expected that
we can distinguish the Higgs signal from backgrounds
by measuring the hadronic ($q\bar{q}$) invariant mass $M_{q\bar{q}}$
distribution of the candidate events, which should have
a distinct peak for the SM Higgs events as we can see from Fig.9.
However, if the Higgs decays into the stops
$h$ $\to$ $\stl\stls$ $\to$ $c\bar{c}\sz1\sz1$, the peak in
$M_{q\bar{q}}$ distribution will appear at
$M_{q\bar{q}}$ $\ll$ $\mh$ because
the large energies and momenta will be carried off by
the neutralino $\sz1$.
As a consequence, the signatures will be hided the large backgrounds
from the $\gamma\gamma$ processes and $Z$ $\to$ $q\bar{q}\ell^+\ell^-$
events for
the neutrino channel $Z$ $\goto$ $hZ^{*}$ $\goto$ $h(\nu\overline{\nu})$
and the lepton channel
$Z$ $\goto$ $hZ^{*}$ $\goto$ $h(\ell^{+}\ell^{-})$, respectively.

Usually we can suppress the $\gamma\gamma$ background
by taking a lower cut for the jet transverse momentum for
the neutrino channel $Z$ $\goto$ $hZ^{*}$ $\goto$ $h(\nu\overline{\nu})$.
Such a selection cut will not be useful when the Higgs decays into the
stops since the jet transverse momentum will become very soft.
So we suppose that the expected number of events
of the neutrino channel reduced considerably.
On the other hand,
the selection cuts on the visible jet energies
are not so essential in the lepton channel
$Z$ $\goto$ $hZ^{*}$ $\goto$ $h(\ell^{+}\ell^{-})$.
The energy and momentum distribution
of the scattered leptons from the virtual
$Z$-boson are irrespective to the decay modes of the Higgs.
In fact, the Higgs mass can be determined by the recoil mass
\begin{equation}
M_{recoil}^{2}=\mz^2-2\mz\left(E_{\ell^+}+E_{\ell^-}\right)
+M_{\ell^+\ell^-}^2,
\end{equation}
instead of the direct measurement of $M_{q\bar{q}}$ in principle.
We suppose, therefore, that
the reduction rates of the events in the lepton channel
are not so large
and that the lower mass bounds on the Higgs from
the lepton channel will be applicable even when
the Higgs decays into the stop pairs.
The Higgs mass bounds
in terms of the lepton channel reported by
the four groups at LEP are slightly different each other
\cite{DELPHI,visZ,LEPh}.
Here we assume conservative bound
\begin{equation}
\mh \nge 55\gev
\end {equation}
and adopt it as a further constraint on our parameter space
($\mu$, $\tanbe$ , $m_t$ ).

\subsection{\it typical parameter sets}

{}From the discussions given above
we find that
the type~{\uppercase\expandafter{\romannumeral 2}} solution
is  more favorable.
In this subsection we present the typical parameter values
of the type~{\uppercase\expandafter{\romannumeral 2}} solution.
Those for the type~{\uppercase\expandafter{\romannumeral 1}} solution
can be found in Ref.\cite{ours}.
Contours of $\mstl$ $=$ $15\gev$
in the ($\mu$, $\tanbe$) plane for $M_{2}$ $=$ $22\gev$
are shown in Fig.10.
Each line corresponds to a contour of
$\mstl=15$GeV for a  fixed $\mt$ value.
We also plot the $\mh=55\gev$ contour and
$\mswl=45\gev$ contour.
In the region denoted by "excluded theoretically"
the false vacuum is realized \cite{susyrho}, i.e.,
\begin{equation}
A^{2}_{t} > 3(m^{2}_{\st_{L}}+m^{2}_{\st_{R}}+m^{2}_{2}),
\end{equation}
where $m^2_2$ $=$
${\widetilde{m}}^{2}_{L_{1}}+\mu^2-3{\widetilde{I}}$.
We see that there is a rather narrow (but finite) range
allowing the existence of the light stop, if the top is
slightly light too, $\mt\nle 95\gev$.
Furthermore, we find that the light stop solution gives inevitably
a light Higgs boson, $\mh\nle 60\gev$.
While we have included the radiative correction in the calculation
of the Higgs mass \cite{higgsmass}, deviations $\delta\mh$ from
the tree level results are not so large, $|\delta\mh|\nle 2\gev$.
The neutral Higgs is standard Higgs like, i.e.,
$\sin(\beta-\alpha)\simeq 1$.

Adopting the bound $\mh\nge 55\gev$,
we can choose three typical parameter sets
(A), (B) and (C), denoted in Fig.10.
Input and output values of the parameters of the sets
(A), (B) and (C) are presented in
Table {\uppercase\expandafter{\romannumeral 1}}.
Interested reader can compare those parameter sets of
the type~{\uppercase\expandafter{\romannumeral 2}} solution
with those of the type~{\uppercase\expandafter{\romannumeral 1}}
solution presented in the previous papers \cite{ours}.
The set (A) [(C)] has the largest [smallest] values of the scalar fermion
masses, the neutral Higgs $h$ mass and the top mass,
and has the smallest [largest] value of the lighter chargino mass.
We find that masses and mixing parameters are severely constrained,
for example,
$\msg\simeq75\gev$, $\mswl\nle50\gev$, $\msl\simeq\msq\simeq300\gev$,
$\tht\simeq0.9$, $\mh\nle60\gev$ and
$\mt\simeq90\gev$.
These parameter sets are characterized by
the rather light top quark and the heavy squarks and sleptons.
To obtain those values we take
$M_2$ $=$ 22GeV and $\sw$ $=$ $0.232$ ($\alpha_{s}=0.113$).
It should be noted that masses of the gluino, the Higgs
and the top depend sensitively on those input parameters.
For example, the allowed ranges are changed to
$\msg\simeq85\gev$, $\mh\nle65\gev$ and $\mt\simeq98\gev$ for
$M_2$ $=$ 24GeV and
$\sw$ $=$ $0.230$ ($\alpha_{s}=0.120$).
Anyway, it seems that the expected top mass
$\mt=90\sim98\gev$ is too small compared to the present
experimental bounds from Tevatron.
We will discuss the top mass bound taking into
account of the stop modes
of the top decay in the next section.

\section{\it Phenomenological implications }

Now we are in position to discuss some consequence of the light stop
scenario in the MSGUT and give strategies to confirm or reject
such a possibility in present and future experiments.
Some numerical results are calculated with the typical parameter sets
(A), (B) and (C) in Table~{\uppercase\expandafter{\romannumeral 1}}.

\subsection{\it top decay}

The existence of the light stop with mass 15 $\sim$ 16GeV will
alter completely decay patterns of some ordinary and SUSY particles
(sparticles).
First we discuss the top decay \cite{Baer,topdecay}.
In our scenario, the top can decay into final states including the stop;
$t\goto$ $\stl\sz1$, $\stl\szs$ and $\stl\gluino$.
Branching ratios of the top for the typical parameter sets are
presented in Table~{\uppercase\expandafter{\romannumeral 2}}.
We find that the stop modes $t\goto\stl\szk , \stl\gluino$ have about
80\% branching ratio and dominate over the
standard mode $t\goto bW^{+}$ $\simeq$ 20\%.
Strategies for the top search at Tevatron
would be forced to change because the leptonic branching ratios of
the top would be reduced by the dominance of the stop modes.
The expected event rates for the ordinary $bW$ signatures will be
reduced, i.e.,
the total cross section
$\sigma(p\bar{p}\to t\bar{t}X\to b\bar{b}W^+W^-X)$
for $\mt\simeq90\gev$ is almost equivalent to that for
$\mt\simeq165\gev$ in the standard model.

\subsection{\it Higgs decay}

Decay patterns of the Higgs particles will be changed too.
The lighter CP-even neutral Higgs decays dominantly into the stop
pair, $h\goto\stl\stls$, owing to the large Yukawa coupling of the top.
In rough estimation, we obtain
\begin{equation}
{\rm BR}(h\goto\stl\stls) \simeq
{\frac{1}{1+{\frac{3\mb^{2}\mh^{2}}{2\mt^{4}}}}} \simeq 1 .
\end{equation}
This fact would change the experimental methods of the Higgs
searches at the present and future collider experiments.
In particular, it could be expected that
the number of events of the neutrino channel
$Z$ $\goto$ $hZ^{*}$ $\goto$ $h(\nu\overline{\nu})$
will be smaller than that of the lepton channel
$Z$ $\goto$ $hZ^{*}$ $\goto$ $h(\ell^{+}\ell^{-})$
in the Higgs $h$ searches at LEP.
This is because the selection cuts on the visible jet energies
in the neutrino channel are large enough to throw away
the soft jets signatures from the Higgs decay $h\to\stl\stls$.
Consequently, when the Higgs decays into the stop pair,
the searches for the lepton channel
$Z$ $\goto$ $hZ^{*}$ $\goto$ $h(\ell^{+}\ell^{-})$
would be more important than the neutrino channel.
More detail analyses of the charged \cite{higgsdecay} and
neutral Higgs bosons \cite{prep} are presented separately.

\subsection{\it sparticle decay}

Now we discuss briefly the light stop impact on the sparticle
decays.
The lightest charged sparticle apart from the stop
is the lighter chargino $\swl$.
The two body stop mode $\swl\goto b\stl$ would dominate over the
conventional three body mode $\swl\goto f\overline{f'}\sz1$.
As a consequence, it would be difficult to use the leptonic
signature in the chargino search at $e^+e^-$ and hadron colliders.
Since the chargino $\swl$, the second neutralino $\szs$,
the neutral Higgs $h$ and
the gluino $\gluino$, whose dominant decay modes are respectively
$\swl\goto b\stl$, $\szs\goto \stl\stls\sz1$, $h\goto \stl\stls$ and
$\gluino\goto \stl\stls\sz1$,
are copiously produced in the other sparticle decays,
many stops would be expected in the final states of the
sparticle production. For example,
$\sll$ $\goto\nu\swl$ $\goto\nu(b\stl)$,
$\slepton$ $\goto\ell\szs$ $\goto\ell(\stl\stls\sz1)$,
$\squark_{L,R}$ $\goto q\gluino$ $\goto q(\stl\stls\sz1)$,
$\sql$ $\goto q'\swl$ $\goto q'(b\stl)$ and
$\zino_{i(i=3,4)}$ $\goto\sz1 h$ $\goto\sz1(\stl\stls)$.

\subsection{\it stop and neutralino searches at LEP}

We can see that
the stop and its relatively light accompaniments,
the gluino $\gluino$, light neutralinos $\zino_{1,2}$, and
neutral Higgs $h$, should be visible at LEP, SLC, HERA and Tevatron.
Especially, LEP could search the allowed region presented in
Figs.8(b) and 10 in terms of the width of $Z$-boson and
the direct stop search.
We can see from Table {\uppercase\expandafter{\romannumeral 1}},
the stop mixing angle $\tht$ is severely
limited as $\tht$ $\simeq$ 0.9 in the allowed range.
As the stop search in terms of $\Delta\Gamma_{Z}$
would be difficult in this case,
the direct search for $e^+e^-\goto\stl\stls$ will be
important \cite{Kobayashi,DELPHI}.
Second, the whole allowed region in Figs.8(b) and 10 can be explored by the
precise measurement of BR($Z\goto vis.$).
In fact, the smallest value of the neutralino contribution
$\sum_{{i,j}\atop{i=j\neq 1}}\Gamma(Z\goto\zino_{i}\zino_{j})/
\Gamma_{Z}^{\rm tot}$ to BR($Z\goto vis.$) is about
$4\times 10^{-7}$.
Here we included the reduction of the ovserved
branching ratio originated from the stop mode in the second
neutralino decay.
Values of branching ratios of the second neutralino
for the typical parameter sets
(A), (B) and (C) are tabulated in
Table {\uppercase\expandafter{\romannumeral 3}}.
Clearly, the lighter chargino, $\mswl\nle 50\gev$, would be visible
at LEP-{\uppercase\expandafter{\romannumeral 2}}.
Moreover, LEP-{\uppercase\expandafter{\romannumeral 2}} may be
able to produce both the top and the stop through a process
$e^+e^-\to t\stl\sz1$.

\subsection{\it gluino and stop searches at Tevatron}

As mentioned before, Tevatron will play a crucial role
in confirming or rejecting the light stop scenario in the MSGUT
with {\it the GUT relation}.
In this case the existence of relatively light gluino,
$\msg$ $=$ $75\sim85\gev$,
with substantially large decay fraction $\gluino\goto\stl\stls\sz1$ is
one of definite prediction.
Values of branching ratios of the gluino for the typical parameter sets
(A), (B) and (C) are tabulated in
Table~{\uppercase\expandafter{\romannumeral 4}}.
The branching ratio of the direct decay mode
BR($\gluino\goto q\overline{q}\sz1$) $=$ $0.4\sim0.5$\% is expected
in the allowed range.
In the gluino search at Tevatron the mixed signature,
$p\overline{p}$ $\goto$ $\gluino\gluino X$ $\goto$
$(\stl\stls\sz1)(q\overline{q}\sz1)X$,
and in turn the two-jets events would be dominant signature.
Of course, the main fraction of the events will be almost invisible
because the resulting charm-jets from
$p\overline{p}$ $\goto$ $\gluino\gluino X$ $\goto$
$(\stl\stls\sz1)(\stl\stls\sz1)X$ $\goto$
$c\bar{c}\sz1\sz1\sz1\sz1\sz1\sz1$
will be too soft to detect.

As has been pointed out by Baer et al. \cite{Baer,Tata} that
the softness of the charm jets makes direct searches for the light stop
difficult at Tevatron.
However, there is a possibility for the detection of the stop ;
the searching for the $\gamma\gamma$ events originated
from the decay of the $s$-wave stoponium
$\sigma_{\stl}=(\stl\stls)$ \cite{stoponium}.

\subsection{\it stop searches at HERA}

The $ep$ collider HERA could search the light stop through
its pair production process
$ep\goto e\stl\stls X$ via boson-gluon fusion \cite{stopbg}.
The total cross section of the process is larger than about
$10 pb$ for $\mstl\nle$20GeV, which is independent on the
mixing angle $\tht$. That is, $\sigma\nge 10 pb$ is expected
for all parameters with $\mstl\nle$20GeV in the allowed range
in Fig.10.
Detail analyses with Monte Carlo studies including
possible dominant background process
$ep\goto ec\cbar X$ can be found in Ref.\cite{sthera}.

\subsection{\it stop searches at SLC}

Polarized initial electron beams at SLC and at any linear $e^+e^-$
colliders will be efficient to reveal the nature of left-right
mixing in the stop sector, in other words, to measure the mixing angle
of stop $\tht$.
In Fig.11 we show the $\rs$ dependence of the left-right asymmetry ;
\begin{equation}
A_{LR}\equiv{\frac{\sigma(e_{L})-\sigma(e_{R})}
                  {\sigma(e_{L})+\sigma(e_{R})}},
\label{asym}
\end{equation}
where $\sigma(e_{L,R})$ $\equiv$
$\sigma(e^{+}e^{-}_{L,R}\goto\stl\stls)$, which are obtained by
\begin{eqnarray}
&&\sigma(e^{+}e^{-}_{L\atop R}\goto\stl\stls) \\
&&={\frac{\pi\alpha^2}{s}}\beta_{\stl}^{3}
\left[{\frac{4}{9}}+{\frac{2}{3}}C_{\stl}(v_{e}\pm a_{e})
{\rm Re}\left({\frac{s}{D_{Z}}}\right)+
{\frac{1}{4}}C_{\stl}^{2}(v_{e}\pm a_{e})^{2}
\left|{\frac{s}{D_{Z}}}\right|^{2}\right](1+\delta_{QCD}),
\end{eqnarray}
where $\beta_{\stl}$ $\equiv$ $\sqrt{1-4\mstl^{2}/s}$,
$D_{Z}$ $\equiv$ $s-\mz^{2}+i\mz\Gamma_{Z}$,
$v_{e}$ $\equiv$ $(-{\frac{1}{2}}+2\sw)/(\sw\cw)$ and
$a_{e}$ $\equiv$ $-1/(2\sw\cw)$.
In the asymmetric combination in Eq.(\ref{asym}) the photon
contribution is cancelled out and $A_{LR}$ is proportional to
$C_{\stl}$  (\ref{c}).
This is the reason for sensitive dependence on $\tht$ of $A_{LR}$
in Fig.11.
Another important property is that $A_{LR}$ is independent on the mass
of stop $\mstl$ as well as on the QCD correction $\delta_{QCD}$
since $\beta_{\stl}^{3}$ and $1+\delta_{QCD}$ disappeared in the
fractional combination of $\sigma$ in $A_{LR}$.
Therefore, this method for measuring $\tht$ will be applicable for
the stop with any mass satisfying $\mstl<\rs/2$.

\section{\it Conclusion}

We have investigated the possibility for
the existence of the light stop
$\mstl=15\sim16\gev$ and the neutralino $\msz1=13\sim14\gev$
in the MSGUT scenario taking into account of the present
experimental bounds on the SUSY parameter space.
We have found that the two types of solution for the RGEs
in the MSGUT,
type~{\uppercase\expandafter{\romannumeral 1}} and
type~{\uppercase\expandafter{\romannumeral 2}},
allowing the existence of the light stop.
It has been found that the type~{\uppercase\expandafter{\romannumeral 2}}
solution is more favorable because these parameter sets seem to
satisfy all constraints settled by the recent collider experiments
as well as by the non-accelerator observations and
the cosmological considerations.
However, the allowed parameter region is rather restricted
and in turn masses
and mixing parameters of
the other SUSY partners as well as masses of the Higgs and
the top are severely constrained.
For example,
$75\gev$ $\nle$ $\msg$ $\nle$ $85\gev$,
$\mswl$ $\nle$ $50\gev$,
$\msl$ $\simeq$ $\msq$ $\simeq$ $300\gev$,
$\tht\simeq0.9$,
$\mh$ $\nle$ $65\gev$ and
$90\gev$ $\nle$ $\mt$ $\nle$ $100\gev$.

Our results can be stated in other words as follows.
In the MSSM the very light stop and the light neutralino
could be allowed only when
the top is relatively light $\mt\nle100\gev$,
the sfermions are heavy $\msf\nle300\gev$ and
the stop mixing angle is in the range
$0.85\nle\tht\nle1.15$.
In this case, the existence of the light stop can compromise with
the negative results of the searches for the top quark, the Higgs
boson and the sparticles at LEP and Tevatron
as well as with the non-observation of the nucleon decay and
the cosmological dark matter constraints.
Then we have found that such a peculiar solution can be realized
in the framework of the MSGUT model.
It is interesting, moreover, that
the mass ratio of the top to the bottom quark
is almost the same as the ratio of the sbottom
to the stop, $\mt/\mb$ $\simeq$ $\msb/\mst$, in our solution.
This may have its origin in a specific model at
the Plank scale such as superstrings.

It should be emphasized that the signatures of
the top and the neutral Higgs will be
changed significantly.
The branching ratio for the standard decay mode of the top
$t\to bW$ will be about 20\% and then
the expected event rates for the ordinary $bW$ signatures will be
reduced, i.e.,
the total cross section
$\sigma(p\bar{p}\to t\bar{t}X\to b\bar{b}W^+W^-X)$
for $\mt\simeq90\gev$ is almost equivalent to that for
$\mt\simeq165\gev$ in the standard model.
For Higgs searches at LEP,
it is expected that the Higgs with mass about $60\gev$
will be discovered more easily in searching for
the lepton channel $Z$ $\goto$ $hZ^{*}$ $\goto$ $h(\ell^{+}\ell^{-})$
than in searching for the neutrino channel
$Z$ $\goto$ $hZ^{*}$ $\goto$ $h(\nu\bar{\nu})$.
Moreover, the light stop and its relatively light accompaniments,
the gluino $\gluino$, the light neutralinos $\zino_{1,2}$, and
the lighter chargino $\wino_1$, should be visible near future
at LEP, HERA and Tevatron.

We have exemplified that if we discover the light stop we will
be able to constrain severely all the SUSY parameters
at the unification scale.
We can conclude that, therefore, the discovery of the stop will bring
us a great physical impact.
Not only will it  be the first signature of the top flavor and
supersymmetry but also it could shed a light on the physics at
the unification scale.

\vskip 20pt

\begin{flushleft}
{\Large{\bf Acknowledgements}}
\end{flushleft}
One of the authors (T.K.) would like to thank R. Enomoto ,
S. Asai and M. Yamaguchi for valuable information
and T. Goto , K. Hagiwara , K. Hidaka , K. Hikasa ,
J. Hisano , M. Nojiri and Y. Okada for helpful comments.
T. K. is also grateful to Y. Chikashige, I. Ito, S. Kitamura
and T. Kobayashi for stimulating discussions.

\vfill\eject

\begin{flushleft}
{\Large{\bf Appendix A}}
\end{flushleft}
In this Appendix we list all renormalization group equations
which are presented in Ref.\cite{Hikasa}.
In the following, dots(\,$\dot{}$\,) denote the derivative {$d/{dt}$}
where $t\equiv(4\pi)^{-1}\ln(\mgut^2/q^2)$.
The coefficients of the one loop $\beta$ function
in the MSGUT are given by
$b_1=33/5$, $b_2=1$ and $b_3=-3$.
The canonical weak hypercharge coupling is defined by
$\alpha'=(3/5)\alpha_1$.
\begin{description}
\item{Gauge couplings}
\begin{eqnarray*}
\dot{\alpha_{i}}&=&-b_{i}\alpha_{i}^{2}.
\end{eqnarray*}
\item{Yukawa Couplings}
\begin{eqnarray*}
\dot{\alpha_{t}}&=&
({\frac{16}{3}}\alpha_{3}+3\alpha_{2}+{\frac{13}{9}}\alpha'-6\alpha_{t})
\alpha_{t}.
\end{eqnarray*}
\item{Higgsino mass $\mu$}
\begin{eqnarray*}
\dot{\mu}&=&
({\frac{3}{2}}\alpha_{2}+{\frac{1}{2}}\alpha'-{\frac{3}{2}}\alpha_{t})
\mu .
\end{eqnarray*}
\item{Gaugino masses}
\begin{eqnarray*}
\dot{M_{i}}&=&-b_{i}\alpha_{i}M_{i}.
\end{eqnarray*}
\item{Scalar masses}
\begin{eqnarray*}
\dot{\mtil}^{2}_{Q_{1,2}}&=&
{\frac{16}{3}}\alpha_{3}M_{3}^{2}
+3\alpha_{2}M_{2}^{2}
+{\frac{1}{9}}\alpha'M_{1}^{2}, \\
\dot{\mtil}^{2}_{Q_{3}}&=&
{\frac{16}{3}}\alpha_{3}M_{3}^{2}
+3\alpha_{2}M_{2}^{2}
+{\frac{1}{9}}\alpha'M_{1}^{2}
-\alpha_{t}(\mssqh+\mssuh+\msh2+A_{t}^{2}), \\
\dot{\mtil}^{2}_{U_{1,2}}&=&
{\frac{16}{3}}\alpha_{3}M_{3}^{2}
+{\frac{16}{9}}\alpha'M_{1}^{2}, \\
\dot{\mtil}^{2}_{D_{1,2,3}}&=&
{\frac{16}{3}}\alpha_{3}M_{3}^{2}
+{\frac{4}{9}}\alpha'M_{1}^{2}, \\
\dot{\mtil}^{2}_{U_{3}}&=&
{\frac{16}{3}}\alpha_{3}M_{3}^{2}
+{\frac{16}{9}}\alpha'M_{1}^{2}
-2\alpha_{t}(\mssqh+\mssuh+\msh2+A_{t}^{2}), \\
\dot{\mtil}^{2}_{L}&=&
3\alpha_{2}M_{2}^{2}+\alpha'M_{1}^{2}, \\
\dot{\mtil}^{2}_{E}&=&
4\alpha'M_{1}^{2}, \\
\dot{\mtil}^{2}_{H_{1}}&=&
3\alpha_{2}M_{2}^{2}+\alpha'M_{1}^{2}, \\
\dot{\mtil}^{2}_{H_{2}}&=&
3\alpha_{2}M_{2}^{2}+\alpha'M_{1}^{2}
-3\alpha_{t}(\mssqh+\mssuh+\msh2+A_{t}^{2}).
\end{eqnarray*}
\item{Trilinear parameters $A$}
\begin{eqnarray*}
\dot{A}_{b}&=&{\frac{16}{3}}\alpha_{3}M_{3}
+3\alpha_{2}M_{2}+{\frac{7}{9}}\alpha'M_{1}-\alpha_{t}A_{t}, \\
\dot{A}_{t}&=&{\frac{16}{3}}\alpha_{3}M_{3}
+3\alpha_{2}M_{2}+{\frac{13}{9}}\alpha'M_{1}-6\alpha_{t}A_{t}.
\end{eqnarray*}
\item{Higgs mixing parameter $B$}
\begin{eqnarray*}
\dot{B}&=&3\alpha_{2}M_{2}+\alpha'M_{1}-3\alpha_{t}A_{t}.
\end{eqnarray*}
\end{description}

\vskip 20pt

\begin{flushleft}
{\Large{\bf Appendix B}}
\end{flushleft}
Here we present the analytical solutions of
all RGEs for the general case $f_i\neq 1$ in the Appendix A.
To get following results we take the boundary conditions
presented in the Sec. 4.
\begin{description}
\item{Gauge couplings}
\begin{eqnarray*}
\alpha_i(t)&=&{\frac{\alpha_{\infty}}{1+b_i\alpha_{\infty}t}}.
\end{eqnarray*}
\item{Gaugino masses}
\begin{eqnarray*}
M_{i}(t)&=&\alpha_{i}(t)f_{i}{\frac{ \mgo }{ \alpha_{\infty} }}
= {\frac{ \mgo {f_{i}} }{ 1+b_{i}{\alpha_{\infty}}t }}.
\end{eqnarray*}
At weak scale, these mass parameters are related as
\begin{eqnarray*}
M_3&=&\msg,  \\
M_{2}&=&{\frac{\alpha}{\alpha_{3}\sw}}{\frac{f_{2}}{f_{3}}}M_3,  \\
M_{1}&=&\frac{5}{3} \tan^2\theta_{W}\frac{f_{1}}{f_{2}}M_{2}.
\end{eqnarray*}
\item{Scalar masses (first and second generation sfermion)}
\begin{eqnarray*}
{{\mtil}^2}(t)&=&m^2_{\infty}+{\mgo}^2{\sum_{i=1}^{3}}{c_i}
{\widetilde{\xi_i}}
\end{eqnarray*}
with
\begin{eqnarray*}
\widetilde{\xi_i}&=&{\frac{1}{2b_i}}{f_i}^2
\left[1-{\frac{1}{(1+b_i{\alpha_{\infty}}t)^2}}\right],
\end{eqnarray*}
where
($c_1$, $c_2$, $c_3$) $=$
($1/15$, $3$, $16/3$) for $\sql$,
($16/15$, $0$, $16/3$) for $\sur$,
($4/15$, $0$, $16/3$) for $\sdr$,
($3/5$, $3$, $0$) for $\sll$ and
($12/5$, $0$, $0$) for $\ser$.
\item{Yukawa  couplings}
\begin{eqnarray*}
\alpha_t(t)&=&{\frac{\alpha_{t\infty}
 \prod_{i=1}^3(1+b_i\alpha_{\infty}t)^{c_i^{\alpha_t}/b_i}}
{1+6\alpha_{t\infty}F (t)}}\\
&=&{\frac{\alpha_{t\infty}\biggl(\frac{\alpha_3\bigl(t\bigr)}
{\alpha_{\infty}}\biggr)^{\frac{16}{9}}
\biggl(\frac{\alpha_2\bigl(t\bigr)}{\alpha_{\infty}}\biggr)^{-3}
\biggl(\frac{\alpha_1\bigl(t\bigr)}
{\alpha_{\infty}}\biggr)^{-\frac{13}{99}}}{1+6\alpha_{t\infty}F(t)}},
\end{eqnarray*}
where
\begin{eqnarray*}
F(t)&=&\int_0^{t}dt' \prod_{i=1}^3(1+b_i\alpha_\infty
t')^{c_i^{\alpha_t}/b_i}=\int_0^t
dt'\left(\frac{\alpha_3(t')}{\alpha_\infty}\right)^{\frac{16}{9}}
\left(\frac{\alpha_2(t')}{\alpha_\infty}\right)^{-3}
\left(\frac{\alpha_1(t')}{\alpha_\infty}\right)^{-\frac{13}{99}}
\end{eqnarray*}
with
$c_3^{\alpha_t}=c_3^{\alpha_b}=16/6 , c_2^{\alpha_t}=c_2^{\alpha_b}=3,
c_1^{\alpha_t}=13/15 $ and $ c_1^{\alpha_b}=7/15$.
\item{Higgsino masses}
$$\mu(t)=\frac{\mu_{\infty}\biggl(\frac{\alpha_2\bigl(t\bigr)}
{\alpha_{\infty}}\biggr)^{-\frac{3}{2}}\biggl
(\frac{\alpha_1\bigl(t\bigr)}{\alpha_{\infty}}\biggr)^
{-\frac{1}{22}}}{[1+6\alpha_{t\infty}F(t)]^{1/4}}$$\\
\item{Trilinear paramaters $A$}
\begin{eqnarray*}
A_e(t)&=&\afo+{\mgo}t\left[3f_2\alpha_2(t)+{\frac{9}{5}}f_1
\alpha_1(t)\right],  \\
A_d(t)&=&\afo+{\mgo}t\left[{\frac{16}{3}}f_3\alpha_3(t)+3f_2\alpha_2(t)
+{\frac{9}{5}}f_1\alpha_1(t)\right], \\
A_u(t)&=&\afo+{\mgo}t\left[{\frac{16}{3}}f_3\alpha_3(t)
+3f_2\alpha_2(t)+{\frac{13}{15}}f_1\alpha_1(t)\right], \\
A_\tau(t)&=&A_e(t),  \\
A_t(t)&=&{\frac{\afo}{1+6\alpha_{t\infty}F(t)}}
+\mgo\left[t\sum_{i=1}^3{f_i}{c_i^{\alpha_t}}
{\alpha_i}(t)-{\frac{6\alpha_{t\infty}}
{1+6\alpha_{t\infty}F(t)}}G(f_i;t)\right], \\
A_b(t)&=&\afo+\mgo t \sum_{i=1}^{3} f_i {c_i}^{\alpha_b} \alpha_i(t)-\int_0^t
 dt'\alpha_t(t')A_t(t')  \\
&=&\afo\left[1-{\frac{\alpha_{t\infty}}{1+6\alpha_{t\infty}F(t)}}\right]
+\mgo\left[t\sum_{i=1}^{3} f_i{c_i}^{\alpha_b}
 \alpha_i(t)-{\frac{\alpha_{t\infty}G(f_i;t)}
{1+6\alpha_{t\infty}F(t)}}\right],
\end{eqnarray*}
where
\begin{eqnarray*}
G(f_i;t)&=&\int_0 ^t dt' \sum_{i=1}^3 f_i c_i^{\alpha_t}
\alpha_i(t')t'F'(t').
\end{eqnarray*}
Here we use the relation
\begin{eqnarray*}
\int_0^t dt'\alpha_t(t')A_t(t')&=&\afo
 \frac{\alpha_{t\infty}F(t)}{1+6\alpha_{t\infty}F(t)}
+\mgo \alpha_{t\infty}\Ktil(t)
\end{eqnarray*}
with
\begin{eqnarray*}
\Ktil(t) &=& \int_0^t dt' \frac{G'(f_i;t')}{1+6\alpha_{t\infty}F(t')}
 -\int_0^tdt'
\frac{6\alpha_{t\infty}F'(t')}{[1+6\alpha_{t\infty}F(t')]^2} G(f_i;t') \\
&=& \frac{G(f_i;t)}{1+6\alpha_{t\infty}F(t)}.
\end{eqnarray*}
\item{Higgs mixing paramater $B$}
\begin{eqnarray*}
B(t)&=&B_\infty+\mgo t\left[3f_2 \alpha_2 (t)
+\frac{3}{5}f_1\alpha_1(t)\right]-3\int_0^t
 dt'\alpha_t(t')A_t(t')\\
&=&B_\infty-3\afo\frac{\alpha_{t\infty}F(t)}{1+6\alpha_{t\infty}F(t)}
+\mgo \left\{
 t\left[3f_2\alpha_2(t)+\frac{3}{5}f_1\alpha_1(t)\right]
-\frac{3\alpha_{t\infty}G(f_i;t)}{1+6\alpha_{t\infty}F(t)} \right\}.
\end{eqnarray*}
\item{Scalar masses (Third Generation and Higgs)} \\
If one negrects the Yukawa couplings except for the top
(valid for $\tanbe\ll\mt/\mb$),
masses for $\staul$, $\staur$, $\nu_{\tau}$ and $\sbr$ are the
same as their lower generation counterparts.
Furthermore, the soft-breaking mass term for
$H_1$, $\widetilde{m}^2_{H_1}$, is equarl to $\widetilde{m}^2_{L}$.
The solutions for the others are found to be
\begin{eqnarray*}
\mtil_{Q_3}^2(t)&=&\mtil_{Q_1}^2(t)-\Itil(t), \\
\mtil_{U_3}^2(t)&=&\mtil_{U_1}^2(t)-2\Itil(t), \\
\mtil_{H_2}^2(t)&=&\mtil_{L_1}^2(t)-3\Itil(t).
\end{eqnarray*}
where
\begin{eqnarray*}
\Itil(t)&=&\frac{\alpha_{t\infty}}
{1+6\alpha_{t\infty}F(t)}\Bigg\{ 3\mfo ^2F(t)\\
&+&\mgo ^2 \sum_{i=1}^3 \frac{c_i^{\Sigma}f_i^2}{2b_i} \int _0 ^t
 dt'F'(t')\left[1-(\frac{\alpha_i(t')}{\alpha_\infty})^2\right]
+\int _0 ^t dt'F'(t')[A_t(t')]^2 \Bigg\}
\end{eqnarray*}
with $c_3^\Sigma=\frac{32}{3}$,$c_2^\Sigma
=6$ {\rm { and }} $c_1^\Sigma=\frac{26}{15}$.
We can write the last integral as folllows ;
\begin{eqnarray*}
\int _0 ^t dt' F'(t')[A_t(t')]^2&=&\afo^2 \frac{F(t)}{1+6
 \alpha_{t\infty}F(t)}+2A_\infty\mgo\Ktil(t)+\mgo^2 \Ktil _2(t),
\end{eqnarray*}
where
\begin{eqnarray*}
\Ktil_2(t)&=&\int _0 ^t dt' F'(t')\left[t'\sum_{i=1}^3 f_i c_i^{\alpha_t}
 \alpha_i(t')-\frac{6 \alpha_{t\infty}G(f_i;t')}
{1+6 \alpha_{t\infty}F(t')}\right]^2.
\end{eqnarray*}
$\Itil(t)$ is a quadratic function of $\mfo , \mgo ,$ and $\afo$ :
\begin{eqnarray*}
\Itil(t)&=&\Itil_{SS}\mfo^2+\Itil_{GG}\mgo^2+\Itil_{GA}
\mgo\afo+\Itil_{AA}\afo^2,
\end{eqnarray*}
where
\begin{eqnarray*}
\Itil_{SS}&=&\frac{3\alpha_{t\infty}F(t)}{1+6\alpha_{t\infty}F(t)}, \\
\Itil_{GG}&=&\frac{\alpha_{t\infty}}{1+6\alpha_{t\infty}F(t)}\int_0^t dt'
 F'(t')\Bigg\{ \sum_{i=1}^3
 \frac{c_i^{\Sigma}f_i^2}{2b_i}\left[1-(\frac{\alpha_i(t')}
{\alpha_\infty})^2\right] \\
&+&\left[t'\sum_{i=1}^3f_ic_i^{\alpha_t}\alpha_i(t')
-\frac{6\alpha_{t\infty}G(f_i;t')}
{1+6\alpha_{t\infty}F(t')}\right]^2\Bigg\}
-6\left[\frac{\alpha_{t\infty}G(f_i;t)}{1+6\alpha_{t \infty}F(t)}
\right]^2, \\
\Itil_{GA}&=&\frac{2 \alpha_{t\infty}}
{[1+6\alpha_{t\infty}F(t)]^2}G(f_i;t),  \\
\Itil_{AA}&=&\frac{\alpha_{t\infty}F(t)}{[1+6\alpha_{t\infty}F(t)]^2}.
\end{eqnarray*}
\end{description}

\vskip 20pt

\begin{flushleft}
{\Large{\bf Appendix C}}
\end{flushleft}
Here we present the results of the MSGUT analyses
without the GUT relation ($f_i \neq 1$).
The calculational scheme is not different from
that in Sec.4 except for $f_3 \neq 1$ .
Now we seek solutions to give the light stop with its mass
$\mstl = 15 \gev$ varing the three paramaters ($\mu ,\tanbe ,m_t$)
for $f_1=f_2=1$ and $f_3 \neq 1$.
In this case we can take
$\msg$ values freely because {\it the GUT relation} is
modified as
\begin{eqnarray*}
\msg = M_{3} = f_3 {\frac{\alpha_{s}}{\alpha}}\sw M_{2}.
\end{eqnarray*}
The typical results are shown in
Table {\uppercase\expandafter{\romannumeral 5}}.
\vskip 20pt

\begin{flushleft}
{\Large{\bf Appendix D}}
\end{flushleft}
In this Appendix we present the formulae for the
widths of the second neutralino decay $\szs\to\stl\stls\sz1$
and the gluino decay $\gluino\to\stl\stls\sz1$.
Formulae for the other modes can be found in Ref.\cite{susydecay}.
The total decay width of the three-body decay
$\Xtil(k)\to\stl(p_2)\stls(p_3)\sz1(p_1)$ can be written as
\begin{eqnarray*}
\Gamma&=&{\frac{1}{(2\pi)^3}}{\frac{1}{32m_{\Xtil}^3}}
\int_{(\msz1+\mstl)^2}^{(m_{\Xtil}-\mstl)^2}
d\mot\int_{(\mor)_{min}}^{(\mor)_{max}}d\mor {\bar{\sum}}_{spin}|M|^{2}, \\
&& \ \ (\mor)_{min\atop max}
=(E_1^*+E_3^*)^2-
\left(\sqrt{E_1^{*2}-\msz1^2}\pm\sqrt{E_3^{*2}-\mstl^2}\right)^2,
\end{eqnarray*}
where
\begin{eqnarray*}
E_1^*&=&{\frac{1}{2m_{12}}}(\mot+\msz1^2-\mstl^2), \\
E_1^*&=&{\frac{1}{2m_{12}}}(m_{\Xtil}^2-\mot-\mstl^2)
\end{eqnarray*}
and $m_{ij}^2\equiv(p_i+p_j)^2$. \\
The matrix elements $M=\sum_{\alpha}M_{\alpha}$
corresponding to the Feynman diagrams Figs.3 and 6
are given by
\begin{eqnarray*}
&&M_1={\frac{2 i e g_s T^a}{\mot-\mt^2}}
\left[\bar{u}(p_1)\left(F_1^L P_R + F_1^R P_L\right)
(\pos + \pts +\mt)\left(\cos{\tht} P_R + \sin{\tht} P_L\right)
u(k)\right], \\
&&M_2={\frac{2 i e g_s T^a}{\mor-\mt^2}}
\left[\bar{u}(p_1)\left(F_1^L P_R + F_1^R P_L\right)
(\pos + \prs +\mt)\left(\cos{\tht} P_R + \sin{\tht} P_L\right)
u(k)\right]
\end{eqnarray*}
for $\gluino\to\stl\stls\sz1$ and
\begin{eqnarray*}
&&M_1={\frac{2 i e^2}{\mot-\mt^2}}
\left[\bar{u}(p_1)\left(F_1^L P_R + F_1^R P_L\right)
(\pos + \pts +\mt)\left(F_2^L P_L + F_2^R P_R\right)
u(k)\right], \\
&&M_2={\frac{2 i e^2}{\mor-\mt^2}}
\left[\bar{u}(p_1)\left(F_1^L P_R + F_1^R P_L\right)
(\pos + \prs +\mt)\left(F_2^L P_L + F_2^R P_R\right)
u(k)\right], \\
&&M_3={\frac{2 i e^2}{\mtr-\mh^2}}
X G_{12} \left[\bar{u}(p_1)u(k)\right]
\end{eqnarray*}
for $\szs\to\stl\stls\sz1$.
The coupling parameters are defined by
\begin{eqnarray*}
&&F_k^L={\frac{\mt N'_{k4}}{2\mw\sin{\tew}\sin{\beta}}}\cos{\tht}
+(e_uN'_{k1}-A_u^RN'_{k2})\sin{\tht} \\
&&F_k^R=(e_uN'_{k1}+A_u^LN'_{k2})\cos{\tht}
-{\frac{\mt N'_{k4}}{2\mw\sin{\tew}\sin{\beta}}}\sin{\tht} \\
&&G_{kl}={\frac{1}{2}}(\sin{\alpha}N_{k3}+\cos{\alpha}N_{k4})
\left({\frac{N_{l2}}{\sin{\tew}}}-{\frac{N_{l1}}{\cos{\tew}}}\right)\\
&&\qquad +{\frac{1}{2}}(\sin{\alpha}N_{l3}+\cos{\alpha}N_{l4})
\left({\frac{N_{k2}}{\sin{\tew}}}-{\frac{N_{k1}}{\cos{\tew}}}\right), \\
&&X={\frac{1}{4}}\Big[\mz\sin{(\alpha+\beta)}
(\cos^2{\tht}A_u^L+\sin^2{\tht}A_u^R)-
{\frac{\mt^2}{\mw\sin{\tew}}}{\frac{\cos{\alpha}}{\sin{\beta}}} \\
&& \qquad +{\frac{\mt}{\mw}}
{\frac{\sin{\tht}\cos{\tht}}{\sin{\tew}\sin{\beta}}}
(\mu\sin{\alpha}-A_t\cos{\alpha})\Big],
\end{eqnarray*}
where $A_u^L=(1/2-e_u\sin^2\tew)/(\cos\tew\sin\tew)$ and
$A_u^R=e_u\tan\tew$.
After calculating the matrix element squared, we get
\begin{eqnarray*}
&&\Gamma(\szs\to\stl\stls\sz1)={\frac{\alpha^2}{8\pi\mszs^3}}
\int d\mot d\mor
\Big[{\frac{T_{11}}{(\mot-\mt^2)^2}}+{\frac{T_{22}}{(\mor-\mt^2)^2}}\\
&& \qquad +{\frac{T_{12}}{(\mot-\mt^2)(\mor-\mt^2)}}
+{\frac{U_{33}}{(\mtr-\mh^2)^2}} \\
&& \qquad +{\frac{U_{13}}{(\mot-\mt^2)(\mtr-\mh^2)}}
+{\frac{U_{23}}{(\mor-\mt^2)(\mtr-\mh^2)}}\Big],
\end{eqnarray*}
where
\begin{eqnarray*}
&&T_{ij}=[(F_1^L F_2^L)^2+(F_1^R F_2^R)^2]T_{ij}^A
+2[(F_1^L)^2+(F_1^R)^2]F_2^L F_2^R T_{ij}^B \\
&& \qquad
+2[(F_2^L)^2+(F_2^R)^2]F_1^L F_1^R T_{ij}^C
+[(F_1^L F_2^R)^2+(F_1^R F_2^L)^2]T_{ij}^D \\
&& \qquad
+4F_1^L F_1^R F_2^L F_2^R T_{ij}^E, \\
&&T_{11}^A=\mot(\mszs^2+\msz1^2-\mor)
         +(\mszs^2-\mstl^2)(\msz1^2-\mstl^2)\\
&&T_{11}^B=\mt\mszs(\mot+\msz1^2-\mstl^2) \\
&&T_{11}^C=\mt\msz1(\mot+\mszs^2-\mstl^2) \\
&&T_{11}^D=\mt^2(\mot+\mor-2\mstl^2) \\
&&T_{11}^E=\msz1\mszs(\mot+\mt^2) \\
&&T_{12}^A=2(\mot\mor+\mszs^2\msz1^2-\mstl^4) \\
&&T_{12}^B=\mt\mszs(\mot+\mor+2\msz1^2-2\mstl^2) \\
&&T_{12}^C=\mt\msz1(\mot+\mor+2\mszs^2-2\mstl^2) \\
&&T_{12}^D=2\mt^2(\mot+\mor-2\mstl^2) \\
&&T_{12}^E=\msz1\mszs(\msz1^2+\mszs^2-2\mstl^2+2\mt^2) \\
&&U_{33}=2X^2G_{12}^2(\mot+\mor-2\mstl^2+2\msz1\mszs) \\
&&U_{13}=2XG_{12}[
(F_1^L F_2^R + F_1^R F_2^L)\mt(\mot+\mor+2\msz1\mszs-2\mstl^2) \\
&& \quad
+(F_1^L F_2^L + F_1^R F_2^R)(\msz1+\mszs)(\mot-\mstl^2+\msz1\mszs)]
\end{eqnarray*}
and
\begin{eqnarray*}
&&T_{22}^{A\sim E} = T_{11}^{A\sim E}(\mot\Leftrightarrow\mor)\\
&&U_{23} = U_{13}(\mot\Leftrightarrow\mor).
\end{eqnarray*}
The formula for $\Gamma(\gluino\to\stl\stls\sz1)$ can be obtained
by following replacements in $\Gamma(\szs\to\stl\stls\sz1)$,
\begin{eqnarray*}
&&\mszs\Rightarrow\msg \\
&&\alpha^2\Rightarrow {\frac{1}{2}}\alpha\alpha_{s}\\
&&U_{ij}=0 \\
&&F_2^L \Rightarrow \cos{\tht} \\
&&F_2^R \Rightarrow \sin{\tht}.
\end{eqnarray*}

\vfill\eject

\vfill\eject

\baselineskip = 18pt plus 1pt
\noindent{\Large{\bf Figure Captions}} \\
\medskip
{\bf Figure 1:} \ \
Excluded regions in ($\mu$, $M_2$) plane by LEP and Tevatron
for $\tanbe=2$.
Regions denoted by A, B, C and D are excluded by
the chargino mass, the visible branching ratio of the $Z$,
the invisible width of the $Z$ and the gluino mass, respectively.
Contours of $\msz1=13\gev$ and $14\gev$ are also depicted.
\label{fig1}
\\
\medskip
{\bf Figure 2:} \ \
Allowed region in ($\mu$, $\tanbe$) plane for $M_{2}=22\gev$
and $24\gev$.
\label{fig2}
 \\
\medskip
{\bf Figure 3:} \ \
Feynman diagrams for the gluino decay
$\gluino\to\stl\stls\sz1$.
\label{fig3}
 \\
\medskip
{\bf Figure 4:} \ \
$\msg$ dependence of branching ratios of gluino.
Sum over quark flavors $q,q' = u, d, c, s$ are taken.
Input parameters are $\tanbe=2.0$, $\mu=-150\gev$, $\mstl=15\gev$,
$\tht=0.9$ and $M_{2}=22\gev$.
We take ($\msq$, $\mt$) $=$ ($2\msg$, $130\gev$) for (a) and
($3.8\msg$, $95\gev$) for (b).
\label{fig4}
 \\
\medskip
{\bf Figure 5:} \ \
Expected number of events from $p\overline{p}$ $\goto$
$\gluino\gluino X$ at CDF.
Input parameters are $\tanbe=2.0$, $\mu=-150\gev$, $\mstl=15\gev$,
$\tht=0.9$ and $M_{2}=22\gev$.
We take ($\msq$, $\mt$) $=$ ($2\msg$, $130\gev$) for
type~{\uppercase\expandafter{\romannumeral 1}} and
($3.8\msg$, $95\gev$) for
type~{\uppercase\expandafter{\romannumeral 2}}.
Straight line corresponds to upper bound settled by
CDF gluino search.
\label{fig5}
 \\
\medskip
{\bf Figure 6:} \ \
Feynman diagrams for the second neutralino decay
$\szs\to\stl\stls\sz1$.
\label{fig6}
 \\
\medskip
{\bf Figure 7:} \ \
$\mt$ dependence of branching ratios of second neutralino.
Input parameters are $\tanbe=2.0$, $\mu=-150\gev$, $\mstl=15\gev$,
$\tht=0.9$, $M_{2}=22\gev$, $\mh=60\gev$, $\alpha=-0.6$ and $\msl=\msq$.
We take ($\msq$, $A_t$) $=$ ($2\msg$, $300\gev$) for (a)
and ($3.8\msg$, $600\gev$) for (b).
\label{fig7}
 \\
\medskip
{\bf Figure 8:} \ \
Excluded regions in ($\mu$, $M_2$) plane ($\tanbe=2$) by LEP and Tevatron
for type~{\uppercase\expandafter{\romannumeral 1}} (a) and
type~{\uppercase\expandafter{\romannumeral 2}} (b) solution.
Definitions of regions are the same as those given in Fig.1.
\label{fig8}
\\
\medskip
{\bf Figure 9:} \ \
Expected number of events from $Z$ $\goto$ $hZ^{*}$ at LEP
for neutrino channel (a) and lepton (muon) channel (b).
Input parameters are $\tanbe=2$, $\mstl=15\gev$, $\msz1=13\gev$,
$\alpha=-0.5$ and $L=93.5pb^{-1}$.Here kinematical
cuts $| \cos \theta_{q} | < $ 0.9 for (a) and
$M_{\ell^+\ell^-}> $ 20 $\gev$ for (b) are adopted.
\label{fig9}
 \\
\medskip
{\bf Figure 10:} \ \
Stop mass contours in ($\mu$, $\tanbe$) plane ($M_{2}=22\gev$)
for fixed $\mt$.
Each line corresponds to contour of
$\mstl=15$GeV for the fixed $\mt$ value.
Points denoted by A, B and C are correspond to
typical parameter sets in the text.
\label{fig10}
 \\
\medskip
{\bf Figure 11:} \ \
Total energy $\rs$ dependence of left-right asymmetry
for the stop production at $e^+e^-$ colliders.
For comparison we also plot $A_{LR}$ for the up-type
quark production.
\label{fig11}
 \\
\vfill\eject

\begin{center}
{{\bf Table {\uppercase\expandafter{\romannumeral 1}}}} \quad
{Typical parameter sets}\\
\end{center}
\begin{center}
\begin{tabular}{|c|rrr|}
\hline
masses in GeV  &     A    &     B    &     C \\
\hline
$M_2$   & $22$     & $22$     &  $22$     \\
$\tanbe$& $2.17$   & $2.0$    &  $2.02$   \\
$\mu$   & $-141$   & $-159$   &  $-146.4$ \\
$\mt$   & $92.5$   & $92.0$   &  $90$     \\
\hline
$\mgo$  & $26.7$   & $ 26.7$  &  $ 26.7$  \\
$\mfo$  & $287.3$  & $276.6$  &  $276.2$  \\
$\afo$  & $692.6$  & $667.1$  &  $672.3$  \\
$\muo$  & $-108.9$ & $-122.8$ &  $-112.7$ \\
\hline
$\mstl$ & $15.0$   & $15.0$   &  $15.0$   \\
$\msth$ & $322.5$  & $310.7$  &  $312.8$  \\
$\tht$  & $0.919$  & $0.923$  &  $0.915$   \\
$\msbl$ & $247.4$  & $237.8$  &  $239.1$  \\
$\msbh$ & $295.8$  & $285.3$  &  $285.0$  \\
$\msul$ & $292.1$  & $281.9$  &  $281.0$  \\
$\msur$ & $293.4$  & $283.0$  &  $282.6$  \\
$\msdl$ & $299.1$  & $288.6$  &  $288.2$  \\
$\msdr$ & $295.4$  & $285.0$  &  $284.6$  \\
$\msll$ & $290.5$  & $279.7$  &  $279.3$  \\
$\mslr$ & $289.7$  & $278.9$  &  $278.5$  \\
$\msn$  & $283.3$  & $272.8$  &  $272.3$  \\
$\mh$   & $58.6$   & $55.6$   &  $54.8$   \\
$\mA$   & $351.7$  & $356.1$  &  $348.0$  \\
$\mH$   & $357.6$  & $362.6$  &  $354.6$  \\
$\mch$  & $357.5$  & $361.8$  &  $354.1$  \\
$\alpha$& $-0.46$  & $-0.49$  &  $-0.49$  \\
$\msz1$ & $13.1$   & $13.1$   &  $13.1$   \\
$\mszs$ & $48.0$   & $48.6$   &  $49.4$   \\
$\mszt$ & $150.5$  & $165.5$  &  $154.1$  \\
$\mszf$ & $178.6$  & $194.2$  &  $183.6$  \\
$\mswl$ & $45.1$   & $45.1$   &  $46.0$   \\
$\mswh$ & $176.4$  & $190.8$  &  $180.5$  \\
$\msg$  & $74.4$   & $74.4$   &  $74.4$   \\
\hline
\end{tabular}
\label{table1}
\end{center}

\vfill\eject

\begin{center}
{{\bf Table {\uppercase\expandafter{\romannumeral 2}}}}\quad
{Branching ratios of top}\\
\end{center}
\begin{center}
\begin{tabular}{|l|ccc|}
\hline
                   &     A    &     B    &     C     \\
\hline
$t\goto\stl\sz1$      & $0.297$  & $0.318$  &  $0.400$  \\
$t\goto\stl\szs$      & $0.231$  & $0.214$  &  $0.286$  \\
$t\goto\stl\gluino$   & $0.261$  & $0.259$  &  $0.128$  \\
$t\goto bW^{+}$       & $0.211$  & $0.209$  &  $0.186$  \\
\hline
\end{tabular}
\label{table2}
\end{center}
\vskip40pt

\begin{center}
{{\bf Table {\uppercase\expandafter{\romannumeral 3}}}}\quad
{Branching ratios of second neutralino}\\
\end{center}
\begin{center}
\begin{tabular}{|l|ccc|}
\hline
                                 &     A    &     B    &     C  \\
\hline
$\szs\goto f\overline{f}\sz1$ & $0.032$  & $0.024$  &  $0.019$  \\
$\szs\goto \stl\stls\sz1$     & $0.968$  & $0.976$  &  $0.981$  \\
\hline
\end{tabular}
\label{table3}
\end{center}
\vskip40pt

\begin{center}
{{\bf Table {\uppercase\expandafter{\romannumeral 4}}}}\quad
{Branching ratios of gluino}\\
\end{center}
\begin{center}
\begin{tabular}{|l|ccc|}
\hline
                                 &     A    &     B    &     C     \\
\hline
$\gluino\to q\overline{q}\sz1$ & $4.4\times10^{-3}$ &
                                 $5.0\times10^{-3}$ &
                                 $4.6\times10^{-3}$  \\
$\gluino\to q\overline{q}\szs$ & $4\times10^{-4}$  &
                                 $4\times10^{-4}$  &
                                 $3\times10^{-4}$  \\
$\gluino\to q\overline{q'}\swl$ & $1.4\times10^{-3}$  &
                                 $1.7\times10^{-3}$  &
                                 $1.3\times10^{-3}$  \\
$\gluino\goto \stl\stls\sz1$     & $0.994$  & $0.993$  &  $0.994$  \\
\hline
\end{tabular}
\label{table4}
\end{center}

\vfill\eject

\begin{center}
{{\bf Table {\uppercase\expandafter{\romannumeral 5}}}}\quad
{Typical parameter sets for $f_3\neq 1$}\\
\end{center}
\begin{center}
\begin{tabular}{|c|rr|rr|}
\hline
masses in GeV & $f_3=0.8$ & & $f_3=2.0$ &            \\
\hline
$M_2$   & $22$     & $22$     &  $22$     &  $22$     \\
$\tanbe$& $2.25$   & $2.05$   &  $2.02$   &  $2.00$   \\
$\mu$   & $-133.4$ & $-140.6$ &  $-156.5$ &  $-126.7$ \\
$\mt$   & $93$     & $90$     &  $95$     &  $90$     \\
\hline
$\mgo$  & $26.7$   & $ 26.7$  &  $ 26.7$  &  $ 26.7$  \\
$\mfo$  & $294.7$  & $280.7$  &  $220.1$  &  $198.4$  \\
$\afo$  & $715.1$  & $688.3$  &  $511.7$  &  $461.8$  \\
$\muo$  & $-103.0$ & $-108.2$ &  $-121.7$ &  $-97.6$  \\
\hline
$\mstl$ & $15.0$   & $15.0$   &  $15.0$   &  $15.0$   \\
$\msth$ & $326.0$  & $313.6$  &  $297.4$  &  $286.3$  \\
$\tht$  & $0.921$  & $0.917$  &  $0.892$  &  $0.874$  \\
$\msbl$ & $250.9$  & $240.2$  &  $220.1$  &  $209.2$  \\
$\msbh$ & $300.5$  & $286.7$  &  $257.3$  &  $239.2$  \\
$\msul$ & $296.7$  & $283.1$  &  $253.4$  &  $234.8$  \\
$\msur$ & $298.0$  & $284.2$  &  $254.7$  &  $236.2$  \\
$\msdl$ & $303.8$  & $290.0$  &  $260.9$  &  $242.9$  \\
$\msdr$ & $300.1$  & $286.3$  &  $256.9$  &  $238.6$  \\
$\msll$ & $297.9$  & $283.8$  &  $215.1$  &  $202.7$  \\
$\mslr$ & $297.1$  & $283.0$  &  $223.0$  &  $201.6$  \\
$\msn$  & $290.6$  & $276.8$  &  $215.1$  &  $193.0$  \\
$\mh$   & $60.2$   & $55.5$   &  $58.1$   &  $55.8$   \\
$\mA$   & $353.1$  & $348.3$  &  $300.3$  &  $260.7$  \\
$\mH$   & $358.8$  & $354.8$  &  $308.2$  &  $270.4$  \\
$\mch$  & $358.9$  & $354.5$  &  $307.4$  &  $270.2$  \\
$\alpha$& $-0.45$  & $-0.48$  &  $-0.50$  &  $-0.52$  \\
$\msz1$ & $13.2$   & $13.2$   &  $13.1$   &  $13.2$   \\
$\mszs$ & $47.6$   & $49.5$   &  $48.6$   &  $51.1$   \\
$\mszt$ & $144.3$  & $149.1$  &  $163.5$  &  $136.2$  \\
$\mszf$ & $171.9$  & $178.6$  &  $192.2$  &  $167.4$  \\
$\mswl$ & $45.0$   & $46.2$   &  $45.1$   &  $48.0$   \\
$\mswh$ & $170.3$  & $175.7$  &  $189.0$  &  $164.3$  \\
$\msg$  & $59.5$   & $59.5$   &  $148.9$  &  $148.9$  \\
\hline
\end{tabular}
\label{table5}
\end{center}
\end{document}